\documentclass[11pt,a4paper]{article}
\usepackage{jheppub}
\usepackage{color}
\usepackage{booktabs}

\newcommand{\fcut}{f_\text{cut}}

\newcommand{\asbar}{{\bar \alpha}_s}

\newcommand{\nfilt}{n_\text{filt}}

\newcommand{\GeV}{\,\mathrm{GeV}}

\newcommand{\as}{\alpha_s}
\newcommand{\order}[1]{{\cal O}\left(#1\right)}
\definecolor{darkgreen}{rgb}{0,0.5,0}
\definecolor{darkblue}{rgb}{0,0,0.7}
\definecolor{darkblue}{rgb}{0.3,0.5,1.0}
\definecolor{darkred}{rgb}{0.5,0,0.0}
\definecolor{darkorange}{rgb}{0.8,0.4,0.0}
\definecolor{brightred}{rgb}{1.0,0,0.0}
\newcommand{\cP}{\mathcal{P}}

\newcommand{\beq}{\begin{equation}}
\newcommand{\eeq}{\end{equation}}
\newcommand{\bea}{\begin{eqnarray}}
\newcommand{\eea}{\end{eqnarray}}
\newcommand{\bdm}{\begin{displaymath}}
\newcommand{\edm}{\end{displaymath}}
\def\as{\alpha_s}

\def\d{\partial}

\def \d{{\rm d} }
\def \d0 {D\O \;}
\def\arctanh{\textrm{arctanh}}
\def\arccoth{\textrm{arccoth}}
\title{Small-radius jets to all orders in QCD}

\author[a]{Mrinal Dasgupta,}
\author[b,c]{Fr\'ed\'eric A. Dreyer,}
\author[d,1]{Gavin P.~Salam,%
\note{On leave from CNRS, UMR 7589, LPTHE, F-75005, Paris, France}}
\author[e]{and Gregory Soyez}
\affiliation[a]{Consortium for
  Fundamental Physics, School of Physics \& Astronomy, University
  of Manchester, Manchester M13 9PL, United Kingdom}
\affiliation[b]{Sorbonne Universit\'es, UPMC Univ Paris 06, UMR 7589,
  LPTHE, F-75005, Paris, France}
\affiliation[c]{CNRS, UMR 7589, LPTHE, F-75005, Paris, France}
\affiliation[d]{CERN, PH-TH, CH-1211 Geneva 23, Switzerland}
\affiliation[e]{IPhT, CEA Saclay, CNRS URA 2306, F-91191 Gif-sur-Yvette, France}

\preprint{
  CERN-PH-TH/2014-222\\
}


\abstract{ 
  As hadron collider physics continues to push the boundaries of
  precision, it becomes increasingly important to have methods for
  predicting properties of jets across a broad range of jet radius
  values $R$, and in particular for small $R$.
  In this paper we resum all leading logarithmic terms, $\alpha_s^n
  \ln^n R^2$, in the limit of small $R$, for a wide variety of
  observables.
  These include the inclusive jet
  spectrum, jet vetoes for Higgs physics and jet substructure tools.
  Some of the quantities that we consider are relevant also for
  heavy-ion collisions.
  Furthermore, we examine and comment on the underlying order-by-order
  convergence of the perturbative series for different $R$ values.
  Our results indicate that small-$R$ effects can be substantial.
  Phenomenological studies will appear in a forthcoming
  companion paper.
}

\begin{document}
\maketitle

%======================================================================
\section{Introduction}

Jets, collimated bunches of energetic hadrons, are widely used at
hadron colliders and lepton colliders as a proxy for hard quarks or
gluons.
Procedures to identify jets group particles into clusters, often based
on their proximity in angle.
Nearly all of the jet algorithms in common use at hadron colliders
follow the example of the seminal work by Sterman and
Weinberg~\cite{Sterman:1977wj} and involve a parameter that determines
how close in angle two particles have to be in order to be clustered
into the same jet.
That parameter is usually referred to as the jet radius, denoted $R$.

From the point of view of perturbative quantum chromodynamics (QCD),
it is natural to choose a jet radius of order $1$ (see e.g.\
Ref.~\cite{KtHH}). 
However in most practical uses of jets at hadron colliders, the jet
radius is taken somewhat smaller: this makes it more straightforward
to resolve multiple jets in events such as top-antitop production,
which can decay to six quarks and in cascade decays of supersymmetric
particles; it also
significantly reduces the contamination of the jet by the underlying
event and multiple simultaneous $pp$ collisions (pileup).
The most common choices for $R$ are in the range
$0.4$--$0.5$~\cite{Aad:2014bia,Chatrchyan:2011ds}, and in some extreme
environments, such as heavy-ion collisions, even smaller values are
used, down to
$R=0.2$~\cite{Ploskon:2009zd,Lai:2009ai,Abelev:2013kqa,Aad:2014wha,Chatrchyan:2014ava}.
Also, one sometimes studies the ratio of the inclusive
jet cross-sections obtained with two different $R$ values
\cite{Abelev:2013fn,Chatrchyan:2014gia,Ploskon:2009zd,Eckweiler:2011mna,Soyez:2011np}. 
Additionally, a number of modern jet tools, such as
filtering~\cite{Butterworth:2008iy} and trimming~\cite{Krohn:2009th},
resolve small subjets within a single moderate-$R$ jet. Yet others
build large-$R$ jets from small-$R$ jets~\cite{Nachman:2014kla}. Many
of these techniques are described in the reviews
\cite{Boost2010,Boost2011,Boost2012,PlehnSpannowsky}.

A problem with small-$R$ jets, which we generically call
``microjets'', is that the correspondence between the jet momentum
and the original parton's momentum is strongly affected by radiation
at angles larger than $R$.
This can degrade momentum measurements with the jets, for example, in
resonance reconstruction.
Furthermore it also affects calculations in perturbative QCD, because
the difference between the parton and jet momenta involves an
expansion whose dominant terms are $\as^n \ln^n R^2$, where $\as$ is
the strong coupling constant:
if $\ln R^2$ is sufficiently large, then the series may no longer
converge, or do so only very slowly.
In such cases in QCD, it is standard to carry out an all-order
resummation. 
Indeed it was argued in Ref.~\cite{Tackmann:2012bt} that this is
a necessity in certain Higgs-boson jet-veto studies. 
Here it is not our intention to argue that all-order resummation of
$\ln R^2$ enhanced terms is an absolute necessity: with a typical
choice of $R=0.4$, $|\ln R^2| \simeq 2$, which is not a genuinely
large number.
However with increasing use of yet smaller $R$ values, it does become
of interest to introduce techniques to carry out small-$R$
resummation.
Furthermore, even for only moderately small $R$ values, a small-$R$
resummation can bring insight and understanding about the origins of
higher-order corrections.

Logarithms of $R$ have been partially resummed before, in a soft
approximation, for jet shapes~\cite{Seymour:1997kj}.
Double logarithms, $(\as \ln R^2 \ln p_{t,\text{cut}})^n$, and first
subleading logarithms have been resummed for jet
multiplicities~\cite{Gerwick:2012fw} above some $p_t$ threshold,
$p_{t,\text{cut}}$. 
Here, using an approach based on angular ordering, we show how to
resum the leading logarithms of $R$ (LL), terms $(\as \ln R^2)^n$, for a wide
range of jet observables, including the inclusive jet spectrum, the
transverse momentum loss from a hard jet, jet veto probabilities, with
results also for filtered and trimmed jets.\footnote{The approach that
  we use has a connection also with the problem of photon isolation in
  small cones, for which the structure of the leading log resummation
  and some phenomenological results were presented in
  Ref.~\cite{Catani:2013oma}.}
In each case, we also include calculations of the coefficients of the
first few orders of the perturbative expansion, which can give insight
into the likely convergence of fixed-order perturbative
calculations.\footnote{Second-order small-R calculations have been
  performed in Refs.~\cite{Alioli:2013hba,vonManteuffel:2013vja}. In
  the case of Ref.~\cite{Alioli:2013hba}, for jet-vetos in Higgs
  production, we give comparisons in
  Appendix~\ref{sec:alioli-walsh}. Ref.~\cite{vonManteuffel:2013vja}
  examined contributions to the resummation of thrust for small-$R$
  jets, which do not appear to be directly comparable to the results
  derived here.}

%======================================================================
\section{Jet Definitions}
\label{sec:jet-defs}

For concreteness we will work with jet algorithms from the generalised
longitudinally-invariant $k_t$ family, which includes the $k_t$~\cite{Kt,KtHH},
Cambridge/Aachen~\cite{Dokshitzer:1997in,Wobisch:1998wt} and
anti-$k_t$~\cite{Cacciari:2008gp} algorithms.
These algorithms introduce a distance measure between every pair of
particles $i$ and $j$
\begin{equation}
  d_{ij} = \min\left(p_{ti}^{2p}, p_{tj}^{2p} \right) \frac{\Delta_{ij}^2}{R^2}
\end{equation}
and a distance measure between each particle and the beam,
\begin{equation}
  d_{iB} = p_{ti}^{2p}\,.
\end{equation}
Here, $p_{ti}$ is the transverse momentum of particle $i$,
$\Delta_{ij}^2 = (y_i - y_j)^2 + (\phi_i - \phi_j)^2$, where $y_i =
\frac12 \ln \frac{E_i+p_{zi}}{E_i-p_{zi}}$ and
$\phi_i$ are respectively the rapidity and azimuth of particle $i$.
The parameter $p$ determines which jet algorithm one is using:
$p=1, 0, -1$ give respectively the $k_t$, Cambridge/Aachen and
anti-$k_t$ algorithm.
$R$ is the jet radius parameter.

The generalised $k_t$ algorithm proceeds by identifying the smallest
of the $d_{ij}$ and $d_{iB}$.
If it is a $d_{ij}$ then particles $i$ and $j$ are merged into a
single new particle and all distances are updated.
If it is a $d_{iB}$, then particle $i$ is called a ``jet'' and removed
from the list of particles.
It is straightforward to see that if a particle $i$ has no other
particle $j$ within a distance $\Delta_{ij} < R$, then $d_{iB}$ will
be smaller than any of the $d_{ij}$,
and the only option for particle $i$ is for it to become a jet.
On the other hand if there is a particle within $\Delta_{ij} < R$,
then those two can merge.
It is in this way that $R$ sets the angular scale on which particles
can recombine into a single jet.

In order to carry out a leading-logarithmic resummation of $(\as \ln
R^2)^n$ terms, we will be considering configurations of particles where
the angles are strongly ordered, e.g.\ $\theta_{12} \gg \theta_{23}
\gg \theta_{34} \gg \ldots$, with $\theta_{ij}$ the angle between
particles $i$ and $j$.
For such configurations, all members of the generalised-$k_t$ family
give identical jets, so we need only carry out a single
resummation. 
The results will be valid also for infrared-safe cone
algorithms such as SISCone~\cite{Salam:2007xv} and also for $e^+e^-$
variants of these algorithms, formulated directly in terms of energies
and angles rather than $p_{ti}$ and $\Delta_{ij}$.
In what follows, for simplicity, we will often simply use angle and
energy variables.

%======================================================================
\section{All-order leading-logarithmic resummation}
\label{sec:gen-func}

The basis of our resummation will be to start with a parton and
consider the emissions from that parton at successively smaller
angular scales.
When we ask questions about (micro)jets with a radius $R$, it is
equivalent at LL order to asking about the set of partons that is
produced by the initial parton $i$ after allowing for all possible
strongly ordered emissions down to angular scale $R$.

It will be convenient to introduce an evolution variable $t$ that
corresponds to the integral over the collinear divergence, weighted
with $\as$ at the appropriate renormalisation scale,
\begin{equation}
  \label{eq:t}
  t(R, p_t) = \int_{R^2}^{1} \frac{d\theta^2}{\theta^2}  \frac{\as(p_t
    \theta)}{2\pi}\,, 
\end{equation}
where $p_t$ here is the transverse momentum of the initial parton.\footnote{
  There is some freedom here on the choice upper limit in angle and on the exact
  scale of $\as$, but these do not matter at the LL accuracy that we are
  targeting.
}
The expansion of $t$ as a power series in $\as$ is
\begin{equation}
  \label{eq:t-power-series}
  t(R,p_t) = \frac{1}{b_0} \ln \frac{1}{1 - \frac{\as(p_t)}{2\pi} b_0  \ln
    \frac{1}{R^2} } 
  = 
  \frac{1}{b_0}\sum_{n=1}^\infty \frac1n \left( \frac{\as(p_t) b_0}{2\pi}
    \ln \frac{1}{R^2}\right)^n\,,
\end{equation}
with $b_0 = \frac{11 C_A - 4 T_R n_f}{6}$.
In a fixed coupling approximation, $t$ is simply $\frac{\as}{2\pi}
\ln\frac1{R^2}$. 

\begin{figure}[t]
  \centering
  \includegraphics[width=0.8\textwidth]{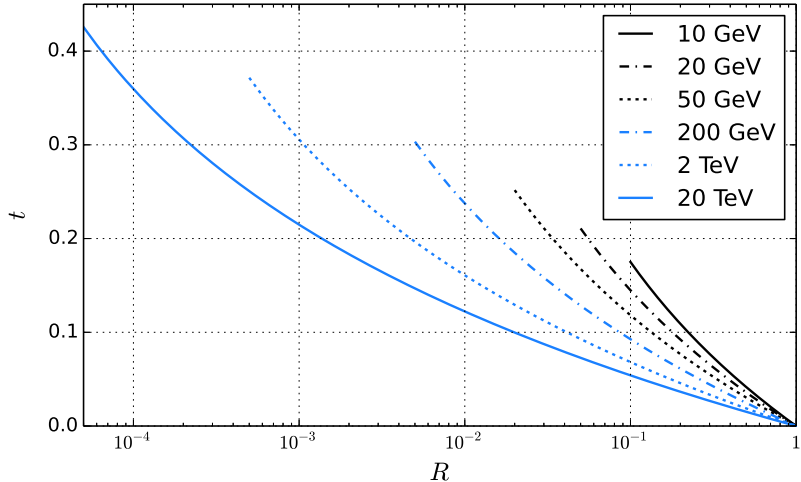}
  \caption{Evolution variable $t$ as a function of angular scale $R$
    for $p_t=10,~20,~50,~200,~2\,000$ and $20\,000 \GeV$. Here $R$ is
    plotted down to $R p_t=1\GeV$.
    For this, and for all other plots that involve a translation
    between $R$ and $t$, we use a one-loop coupling with 5
    flavours such that $\as(M_Z)=0.1184$.
    \label{fig:evolution}
  }
\end{figure}

The dependence of $t$ on the angular scale $R$ is shown in
figure~\ref{fig:evolution} for a range of $p_t$ values.  
The evolution variable is plotted over a range of $t$ such that $R
p_t\ge 1\GeV$, and we see that
in most cases, typical values for $t$ are well below $t\sim0.4$, which
we will therefore take as an upper limit for the rest of this article.
Two reference points that we will use throughout this article are
$R=0.4$ and $R=0.2$, which correspond to $t \simeq 0.041$ and $t\simeq0.077$
respectively for $p_t = 50\GeV$.

To understand the structure of emissions as one evolves in angle, we
will make use of angular ordering~\cite{Dokshitzer:1991wu,Ellis:1991qj}.
This tells us that if an emission $i$ splits to $j$ and $k$ with
opening angle $\theta_{jk}$, then any subsequent emission $\ell$ from
$j$ on an angular scale $\theta_{jl} \ll \theta_{jk}$ is driven purely
by the collinear divergence around $j$ and is independent of the
properties of $k$ (and similarly with $j \leftrightarrow
k$).\footnote{One caveat with angular ordering arises if one looks at
  \emph{azimuthal} correlations. E.g., consider three particles such
  that $\theta_{23} \ll \theta_{12} \ll 1$, and define $\phi_{32}$ as
  the azimuthal angle of particle 3 around particle 2; then the
  distribution in $\phi_{32}$ is not necessarily uniform. In our study
  here, we will not be considering any observables that depend on the
  azimuthal angles, and so this issue does not need to be taken into
  account.}
This will make it relatively straightforward to write an evolution
equation that will encode the full set of potential emission
configurations from an initial hard parton.

To do so, it will be helpful to introduce a generating functional
(cf.\ textbooks such as~\cite{Dokshitzer:1991wu,Ellis:1991qj}).
Suppose that on some angular scale, defined by a $t$-value of $t_1$,
we have a quark with momentum $x p_t$.
We will define a generating functional $Q(x,t_1,t_2)$ that encodes the
parton, or equivalently, microjet, content that one would observe if
one now resolved that parton on an angular scale defined by $t_2 >
t_1$.
The generating functional $Q(x,t_1,t_2)$ satisfies the condition
\begin{equation}
  Q(x,t,t) = q(x)\,,
\end{equation}
where the $q(x)$ term indicates a $100\%$ probability of finding just
a quark with momentum $x p_t$.
To first order in an expansion in $t_2-t_1$ (or equivalently order $\as$),
\begin{equation}
  \label{eq:Qgenfunc_example}
  Q(x,t_1,t_2) = q(x) + (t_2 - t_1) \int_0^1 dz p_{qq}(z) [q(z x)\, g((1-z)x) -
  q(x)] + \order{(t_2-t_1)^2}\,,
\end{equation}
where the real $q \to qg$ splitting function is $p_{qq}(z)= C_F \left(
  \frac{1+z^2}{1-z}\right)$.
Eq.~(\ref{eq:Qgenfunc_example}) indicates that in addition to the
state with just a quark, $q(x)$, at order $t$ there can also be states
with both a quark and a gluon carrying respectively momentum $z x p_t$ and
$(1-z) x p_t$; this is represented by $q(z x)\,
g((1-z)x)$.\footnote{It is important to remember, therefore, that
  $q(x)$ and $g(x)$ are therefore \emph{not} parton distributions.}
As well as a generating functional $Q$ from an initial quark, we have
one from an initial gluon, $G(x,t_1,t_2)$, with the property $G(x,t,t)
= g(x)$.

For concreteness, the mean numbers of quark and gluon microjets of
momentum $z p_t$, on an angular scale defined by $t$, produced from a quark of momentum $p_t$ are
respectively
\begin{equation}
  \label{eq:genfunc-to-number}
  \frac{dn_{q(z)}(t)}{dz} = \left.\frac{\delta Q(1,0,t)}{\delta q(z)}\right|_{\forall q(x)=1, g(x)=1}\,,\qquad
  \frac{dn_{g(z)}(t)}{dz} = \left.\frac{\delta Q(1,0,t)}{\delta g(z)}\right|_{\forall q(x)=1, g(x)=1}\,,
\end{equation}
where $\delta q(z)$ indicates a functional derivative.
For $t=0$, Eq.~(\ref{eq:genfunc-to-number}) consistently gives the
expected result.

%----------------------------------------------------------------------
\subsection{Generating functional evolution equation}

Let us now formulate an evolution equation for the generating
functionals.
We first consider how to relate a quark generating function at initial scale
$0$ to one at an infinitesimal initial scale $\delta_t$:
\begin{multline}
  \label{eq:Q-difference-equation}
  Q(x, 0, t) = Q(x,\delta_t, t) \left(1 - \delta_t \int dz \, p_{qq}(z)
  \right)
  \, + \\
  + \, \delta_t \int dz \, p_{qq}(z)  \bigg[ Q(zx, \delta_t, t)
      G((1-z)x, \delta_t, t) \bigg]\,.
\end{multline}
The term on the first line involves the probability, in large round
brackets, that the initial quark does not branch between scales $0$
and $\delta_t$, so that the partonic content is given by that of a
quark evolving from $\delta_t$ to $t$.
The term on the second line involves the probability that there was a
$q \to qg$ branching in the interval $0$ to $\delta_t$, where the
quark and gluon take fractions $z$ and $1-z$ of the original quark's
momentum and the partonic content is now the combined content of the
quark and the gluon, both evolving from $\delta_t$ to $t$ (this is
represented by the product of generating functionals).
At LL accuracy the generating functionals depend only on the
difference of $t$ values, so we may replace $Q(x,\delta_t, t) = Q(x,0, t
- \delta_t)$.
It is then straightforward to rewrite
Eq.~(\ref{eq:Q-difference-equation}) as a differential equation in
$t$, 
\begin{equation}
  \label{eq:diff-gen-func-Q}
    \frac{dQ(x,t)}{dt} = \int dz \, p_{qq}(z) \left[ Q(zx,t)\,
      G((1-z)x,t) - Q(x,t)\right]\,,
\end{equation}
where we have introduced the shorthand notation $Q(x, t)
\equiv Q(x,0, t)$.
One may proceed in a similar manner for gluons, giving
\begin{multline}
  \label{eq:diff-gen-func-G}
  \frac{dG(x,t)}{dt} = \int dz \, p_{gg}(z) \left[ G(zx,t) G((1-z)x,t)
    - G(x,t)\right] \; +
  \\  
  + \int dz \, p_{qg}(z) \left[
    Q(zx,t) Q((1-z)x,t) - G(x,t)\right]\,.
\end{multline}
where the two further real splitting functions are
\begin{subequations}
  \begin{align}
    %p_{qq}(z) &= C_F \left( \frac{1+z^2}{1-z}\right),\\
    p_{gg}(z) &= 2C_A \left(\frac{z}{1-z} +
      \frac{1}{2}z(1-z)\right),\\
    p_{qg}(z) &= n_f T_R (z^2 + (1-z)^2)\,.
  \end{align}
\end{subequations}
Exploiting the $z \leftrightarrow (1-z)$ symmetry of
Eq.~(\ref{eq:diff-gen-func-G}), we have written $p_{gg}(z)$ such that
it has a divergence only for $z \to 1$.
It will also be convenient to have defined the standard leading-logarithmic
splitting functions including the virtual terms, $P_{qq}(z) =
p_{qq}(z)_+$, $P_{gq}(z) = p_{qq}(1-z)$, $P_{gg}(z) = p_{gg}(z)_+ + p_{gg}(1-z) - \frac{2}{3}n_f
T_R\delta(1-z)$ and $P_{qg}(z) = 2 p_{qg}(z)$ (we sum over quarks and
anti-quarks),
with the usual definition of the plus prescription. 

While
Eqs.~(\ref{eq:diff-gen-func-Q}) and (\ref{eq:diff-gen-func-G}) have
been obtained by introducing an infinitesimal step of evolution at the
beginning of the branching process, it is also possible to write an
equation based on the addition of an infinitesimal step of evolution
at the end of the branching.
For a generating functional $F(t)$ that represents the evolution from
any generic initial condition (i.e. not necessarily a single quark or
a single gluon), the resulting equation reads
\begin{multline}
  \label{eq:diff-gen-func-generic-post}
  \frac{dF(t)}{dt} = \int dx\, dz 
  \left\{
    \frac{\delta F(t)}{\delta q(x)}
    p_{qq}(z) \left[q(z x) g((1-z) x) - q(x) \right]\right.
    + \\ +
    \left. 
      \frac{\delta F(t)}{\delta g(x)}\left[
        p_{gg}(z) (g(zx) g((1-z)x) - g(x))
        + p_{qg}(z) (q(zx) q((1-z)x) - g(x))
      \right]
    \right\}\,.
\end{multline}
The logic of this equation is that for each possible momentum fraction
$x$, one considers all ways of extracting a quark or a gluon with that
momentum fraction, $\delta_{q(x)} F$ or $\delta_{g(x)} F$, and then
integrates over all allowed splittings.
Eqs.~(\ref{eq:diff-gen-func-Q}) and (\ref{eq:diff-gen-func-G}) and
Eq.~(\ref{eq:diff-gen-func-generic-post}) are equivalent, and can be
derived from a common starting point.
Depending on the context, one or the other may be more convenient.

%----------------------------------------------------------------------
\subsection{Fixed-order expansions for the generating functionals}

It is straightforward to solve the coupled pair of equations
(\ref{eq:diff-gen-func-Q}) and (\ref{eq:diff-gen-func-G}) order by
order as a power expansion in $t$.
Writing
\begin{equation}
  Q(x,t) = \sum_n \frac{t^n}{n!} Q_n(x)\,,\qquad\qquad
  G(x,t) = \sum_n \frac{t^n}{n!} G_n(x)\,,
\end{equation}
and making use of the fact that it is sufficient to know just the
result for $x=1$, we have
\begin{subequations}
\begin{align}
  Q_0(1) &= q(1)\,,
  \\
  Q_1(1) &= \int dz \, p_{qq}(z) \left[  q(z)\, g(1-z) - q(1) \right]\,,
   \\
   Q_2(1) &= \int dz dz' \,
   p_{qq}(z) \Big[
   p_{gg}(z')\, \Big( q(z) \, g((1-z) (1-z')) \, g(z'(1-z))-g(1-z)\,  q(z)\Big)+
   \nonumber\\
   &\qquad\qquad
   +p_{qq}(z') \,\Big(g(1-z)\,( g(z(1- z')) \,q(z z')-q(z))
   -g(1-z')\, q(z')+q(1)\Big)+
   \nonumber\\
   &\qquad\qquad
   +p_{qg}(z')\, \Big(q(z)\,q((1-z)(1-z')) \,q((1-z) z')-g(1-z)\, q(z)\Big)\Big]\,,
\end{align}
for the quark case, and
\begin{align}
  G_0(1) &= g(1)\,,
  \\
  G_1(1) &= \int dz \Big[p_{gg}(z) \,(g(z)\, g(1-z) - g(1)) +
  p_{qg}(z)\, ( q(z)\,q(1-z) - g(1) ) \Big]\,,
   \\
   G_2(1) &= \int dz dz' \Big[
   p_{gg}(z) \,p_{gg}(z') \,\Big(g(1-z) \,g(z z') \,g(z(1-z'))-2 g(1-z)\, g(z)+
   \nonumber\\
   &\qquad\qquad\qquad
   +g(z)\, g((1-z) (1-z'))\, g(z'(1-z)) - g(1-z')\, g(z')+g(1)\Big)+
   \nonumber\\
   &\qquad
   +p_{gg}(z')\, p_{qg}(z)\, \Big(g(1)-g(1-z')\, g(z')\Big)+p_{qg}(z)\,
   p_{qg}(z')\, \Big(g(1)-q(1-z')\, q(z')\Big)+
   \nonumber\\
   &\qquad
   +p_{gg}(z)\, p_{qg}(z')\, \Big(g(1-z)\, q(z z')\, q(z(1-z'))-2 g(1-z)\, g(z)+
   \nonumber\\
   &\qquad\qquad\qquad
   +g(z)\, q((1-z) (1-z'))\, q((1-z) z')-q(1-z')\, q(z')+g(1)\Big)+
   \nonumber\\
   &\qquad
   +p_{qg}(z)\, p_{qq}(z')\, \Big(q(1-z)\, g(z-z z')\, q(z z')-2 q(1-z)\, q(z)
   \nonumber\\
   &\qquad\qquad\qquad
   +q(z)\, g((1-z) (1-z'))\, q((1-z)z')\Big)\Big]\,,
\end{align}
\end{subequations}
for the gluon case.
These, and corresponding higher-order expansions, will be used to
obtain the first few orders of the series in $t$ for a range of
observables below.
They can be used both analytically and numerically, by Monte Carlo
integration over the $z$, $z'$, with each term in the integrand
corresponding to a specific partonic configuration.

%----------------------------------------------------------------------
\subsection{All-order reformulation}

Eqs.~(\ref{eq:diff-gen-func-Q}) and (\ref{eq:diff-gen-func-G}) can
be equivalently stated as integral equations
\begin{equation}
  \label{eq:int-gen-func-Q}
  Q(x,t) = \Delta_q(t) Q(x,0) + \int_0^t dt' \Delta_q(t-t') \, 
  \int_\epsilon^{1-\epsilon} dz\, p_{qq}(z) \,Q(zx,t')
  G((1-z)x,t') \, 
\end{equation}
and
\begin{multline}
  \label{eq:int-gen-func-G}
  G(x,t) = \Delta_g(t) G(x,0) + \int_0^t dt' \Delta_g(t-t')
  \int_\epsilon^{1-\epsilon} 
  dz \big[ 
    p_{gg}(z) G(zx,t) G((1-z)x,t') \,+ \\
    p_{qg}(z) Q(zx,t) Q((1-z)x,t')
  \big]
\end{multline}
where we have introduced Sudakov-like form factors:
\begin{align}
  \label{eq:delta-Q}
  \Delta_q(t) &= \exp\left( - t \int_{\epsilon}^{1-\epsilon} dz\, p_{qq}(z)\right),\\
  \label{eq:delta-G}
  \Delta_g(t) &= \exp\left( - t \int_{\epsilon}^{1-\epsilon} dz\,
    (p_{gg}(z) + p_{qg}(z))\right).
\end{align}
The $\epsilon$ cutoffs serve to regularise the divergences in the
splitting functions. 
In the limit $\epsilon \to 0$, the results for $Q(x,t)$ and $G(x,t)$
are independent of $\epsilon$.

The above expressions are suitable for Monte Carlo implementation as a
recursive sequence of splittings, with the Sudakov-like
$\Delta_{q/g}(t)$ factors acting as no-splitting probability
distributions.
We have used such a Monte Carlo implementation, which generates
explicit partonic configurations, for the all-order results
discussed below.\footnote{By ``partonic configuration'', we don't mean
  full 4-vector information, but instead a $z$ momentum fraction for
  each parton and a flavour label, quark or gluon.}
One can similarly reformulate
Eq.~(\ref{eq:diff-gen-func-generic-post}), for which we again have a
Monte Carlo implementation.
It gives identical results.

For numerical purposes we usually take $\epsilon = 10^{-3}$, which we
find is sufficient in order to obtain percent-level accuracy.

Before continuing to the results, it is perhaps worth commenting
on the relation between what we are calculating here and what is
contained in parton-shower Monte Carlo programs. 
We have used an angle as our ordering variable; alternative variables
used in some showers, such as relative transverse momentum or
virtuality differ just by factors of $z$ and/or $1-z$ (and possibly an
overall dimensionful constant).
Because we only consider finite values of $z$ (neither arbitrarily
small, nor arbitrarily close to $1$), the impact of a factor of $z$ or
$1-z$ in the choice of ordering variable is relevant only for
terms beyond LL accuracy.
Thus all parton showers should contain the small-$R$ leading
logarithms that we are resumming here.
One of the main differences between a parton shower and our
calculation (apart from the much wider applicability of a shower) is
that we can isolate a specific physical contribution, making it
possible to obtain analytic results (e.g.\ for the expansion in powers
of $\as$), physical insight and to straightforwardly combine results
with other calculations.

%======================================================================
\section{Results}

In this section we will show illustrative results for a few key
observables of current relevance.
The methods that we use can however be applied more generally.

%----------------------------------------------------------------------
\subsection{Inclusive microjet observables}
\label{sec:incl-microjet}

The most basic collider jet observable is the inclusive jet spectrum,
measured in the past years for example at
HERA~\cite{Aktas:2007aa,Abramowicz:2010ke},
RHIC~\cite{Ploskon:2009zd}, the
Tevatron~\cite{Aaltonen:2008eq,Abazov:2011vi} and
LHC~\cite{Abelev:2013fn,Aad:2014vwa,Chatrchyan:2012bja}.
It has been the subject of many phenomenological studies and
calculations and is of considerable importance notably for
constraining parton distribution functions.
See for example recent progress in NNLO jet
predictions~\cite{Currie:2013dwa,Boughezal:2013uia} and threshold
resummation~\cite{deFlorian:2013qia} and references therein.

Let us introduce the inclusive microjet fragmentation function: given
a parton of flavour $i$, $f^\text{incl}_{j/i}(z,t)$ is the inclusive
distribution of microjets of flavour $j$, at an angular scale defined
by $t$, carrying a fraction $z$ of the parton's moment.
Momentum conservation ensures that
\begin{equation}
  \label{eq:momentum-sum-rule-incl}
  \sum_{j} \int dz z f^{\text{incl}}_{j/i}(z,t) = 1\,.
\end{equation}
In terms of the quantities introduced in section~\ref{sec:gen-func},
$f^{\text{incl}}_{q/q}(z,t)$ is for example nothing other than
$dn_{q(z)}(t)/dz$ as obtained from the $Q(1,t)$ generating functional.

The inclusive microjet fragmentation function trivially satisfies
a DGLAP-style equation
\begin{equation}
  \label{eq:dglap-for-inclusive}
  \frac{df^{\text{incl}}_{j/i}(z,t)}{dt} = 
  \sum_{k} \int_z^1 \frac{dz'}{z'} \,
  P_{jk}(z')\,
  f^{\text{incl}}_{k/i}(z/z',t)\,,
\end{equation}
with an initial condition
\begin{equation}
  \label{eq:dglap-initial-condition}
  f^{\text{incl}}_{j/i}(z,0) = \delta(1-z) \delta_{ji}\,.
\end{equation}
The inclusive microjet spectrum is particularly simple in that one
does not need to make use of the full structure of the generating
functional in order to obtain it.\footnote{Though if one does wish to
  derive it from the generating functional, it is easiest to do so
  using Eq.~(\ref{eq:diff-gen-func-generic-post}).}
Note that a similar result is a part also of the small-$R$ resummation of
the fragmentation contribution to isolated photon production
considered in Ref.~\cite{Catani:2013oma}.

The solution to Eq.~(\ref{eq:dglap-for-inclusive}) can be obtained
using minor adaptations of standard DGLAP evolution codes, e.g.\
QCDNUM~\cite{Botje:2010ay}, QCD-Pegasus~\cite{Vogt:2004ns},
HOPPET~\cite{Salam:2008qg} or
Apfel~\cite{Bertone:2013vaa}.\footnote{An evolution to an angular
  scale defined by $t$ is most straightforwardly mapped to a
  leading-logarithmic DIS evolution, using a $\delta(1-x)$ initial
  condition at some scale $Q_0$ and evolving to a higher scale $Q_1$
  chosen such that $\int_{Q_0^2}^{Q_1^2} \frac{dQ^2}{Q^2}
  \frac{\as(Q^2)}{2\pi} = t$.
  We have explicitly done this with HOPPET and verified that the
  results coincide with those from our Monte Carlo based solution.
}
Alternatively one can use the Monte Carlo solution for the generating
functional outlined in section~\ref{sec:gen-func}, which is the choice
we have made here.

\begin{figure}[t]
  \centering
  \includegraphics[width=0.48\textwidth]{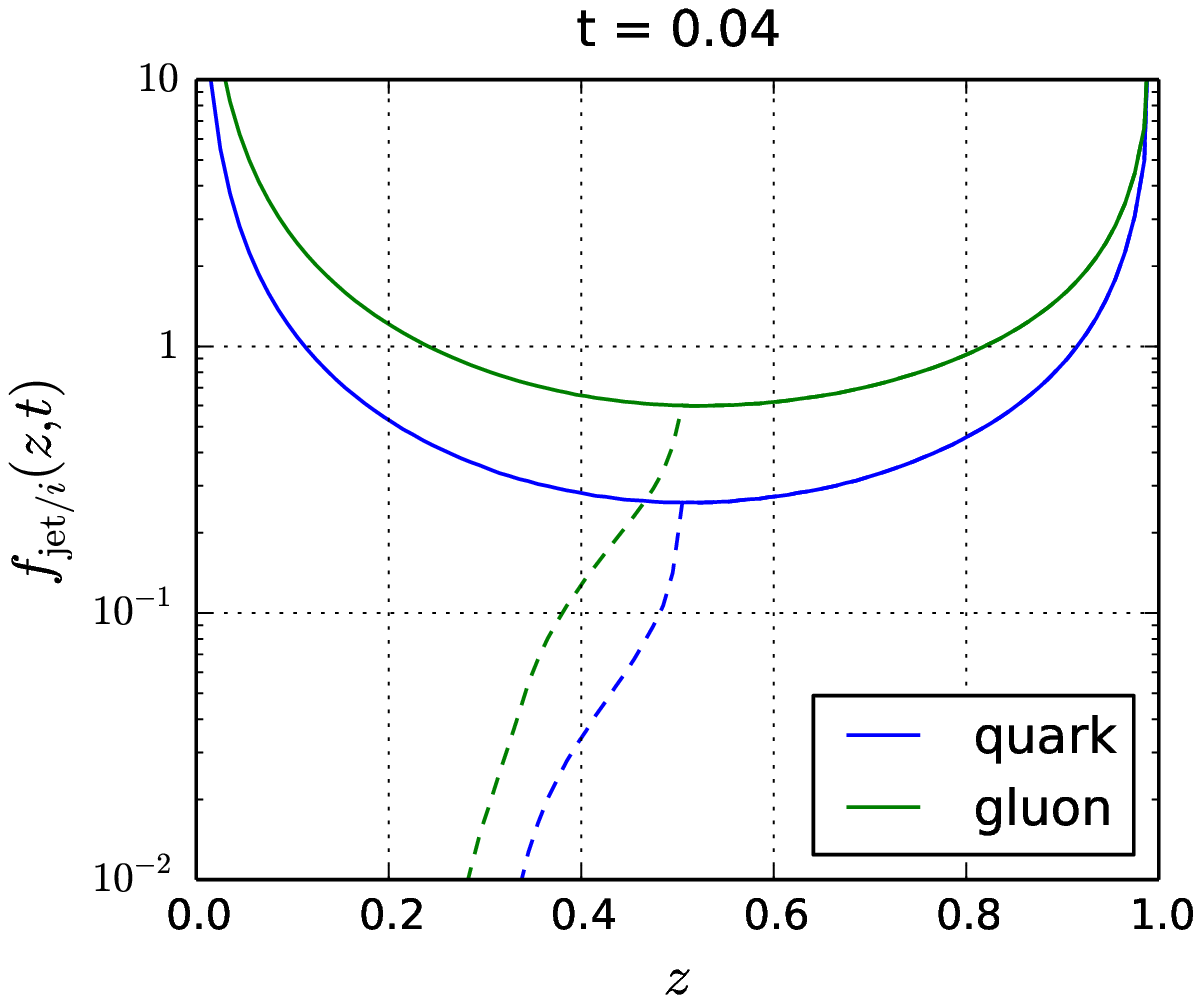}
  \includegraphics[width=0.48\textwidth]{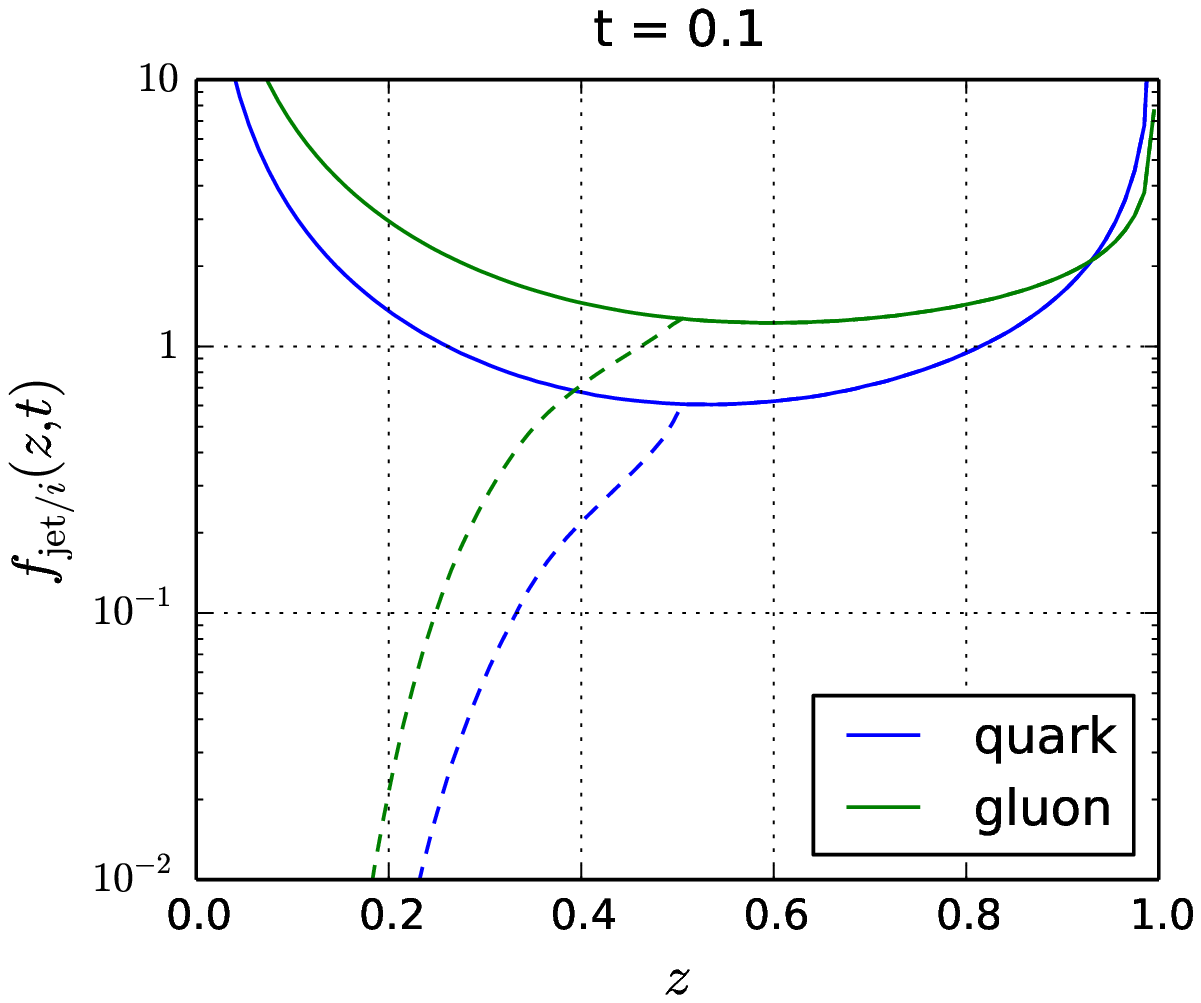}
  \includegraphics[width=0.48\textwidth]{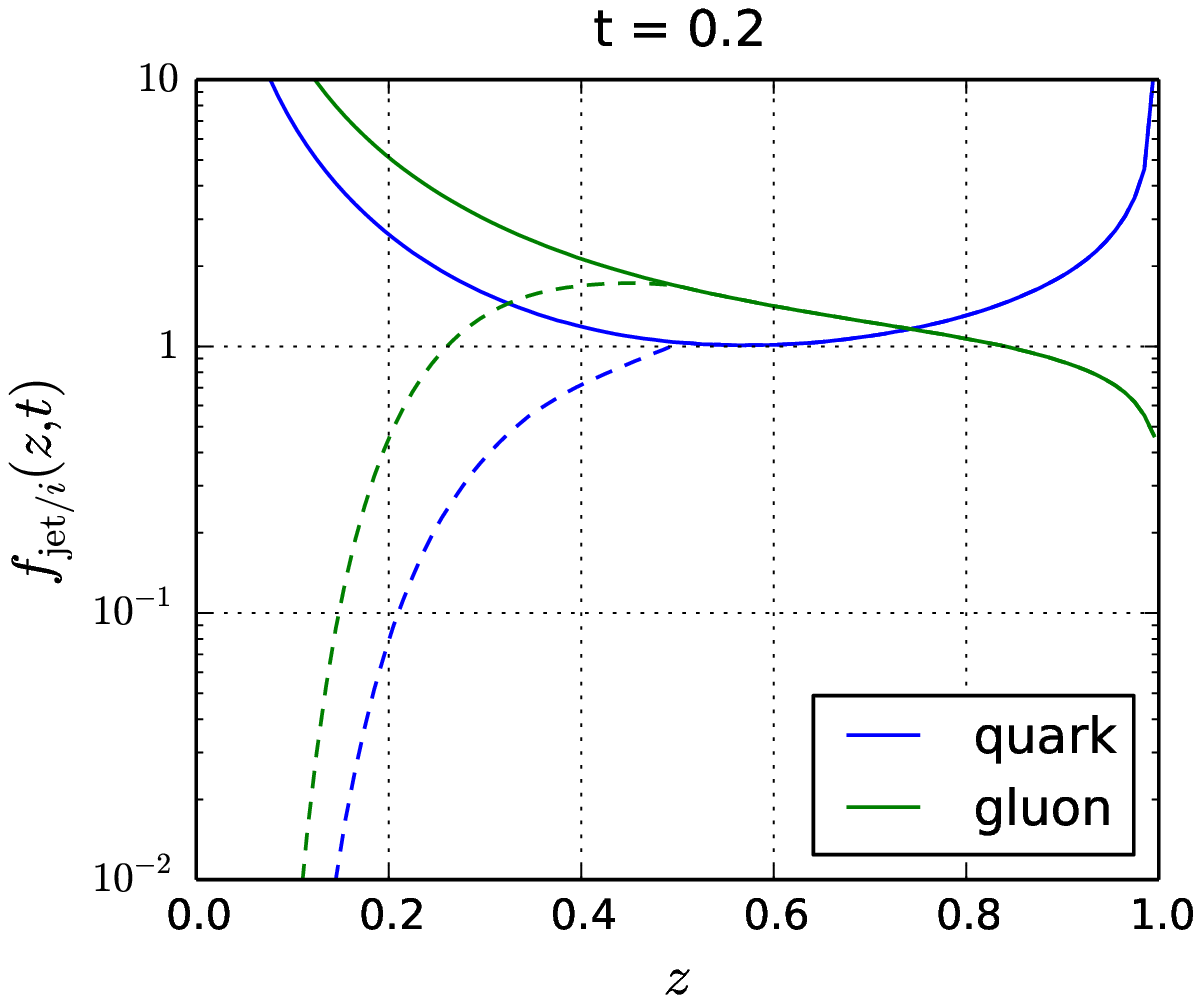}
  \includegraphics[width=0.48\textwidth]{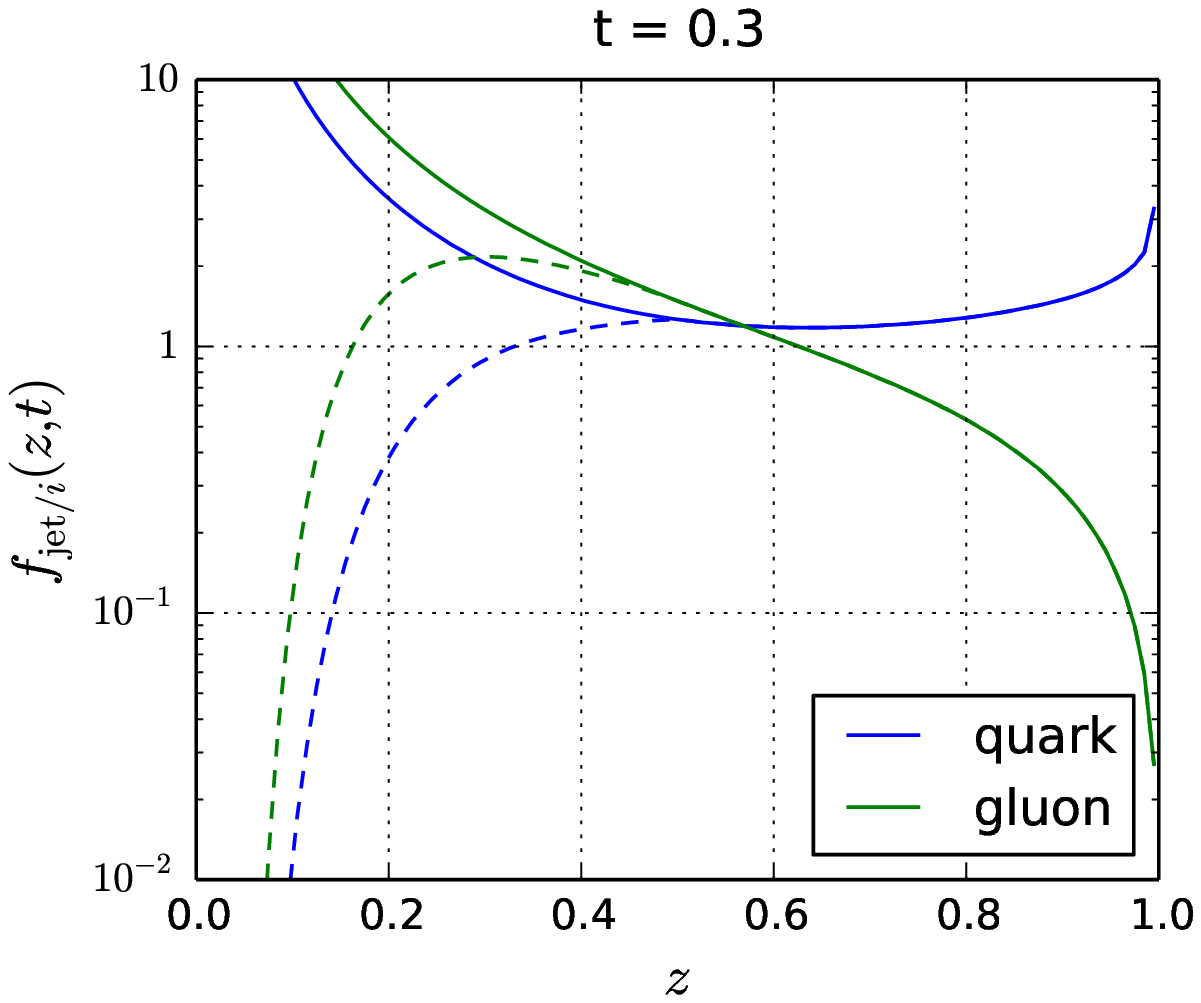}
  \caption{The solid lines show the inclusive microjet spectrum  for
    initial quarks (blue) and initial gluons (green) using LL
    resummation of $\ln R^2$ enhanced terms.
    The dashed lines show the spectrum of the hardest
    microjet.
    They differ from the solid lines only for $z<\frac12$.
    The four panes correspond to $t=0.04,~0.1,~0.2$ and 
    $0.3$.
    \label{fig:fragfct}
  }
\end{figure}

The behaviour of the inclusive microjet fragmentation function when
evolving to smaller angular scales is shown in Fig.~\ref{fig:fragfct}
for $t=0.04,\,0.1,\,0.2$ and $0.3$.
Here, and for all subsequent numerical results, we use $n_f = 5$.
The solid lines represent the inclusive microjet fragmentation
function, in blue for an initial quark and in green for an initial
gluon.
The fragmentation functions are summed over the flavour of the microjets.
One feature of the plots is a peak near $z=1$, showing the presence of
the original parton with an almost unchanged momentum.
As $t$ increases, that peak disappears, and does so more quickly for
initial gluons than for initial quarks.
Away from $z=1$, the fragmentation function for gluons is larger than
that for quarks, as is to be expected given the larger colour factor.
Finally at small $z$, there is a second peak, associated with
production of multiple soft gluon microjets. 
The peak regions do not include resummation of logarithms of $1-z$ for
$z$ near $1$, nor those of $z$ for small $z$. 
The resummation of double logarithms $\as \ln R^2 \ln z$ (and the first
tower of subleading terms) was discussed in
Ref.~\cite{Gerwick:2012fw}.

To examine the impact on a physical observable such as the inclusive
jet spectrum in hadron collisions, it is necessary to convolute the
inclusive microjet fragmentation function with the inclusive partonic
spectrum from hard $2\to2$ scattering.
Let us suppose the partonic spectrum for parton type $i$ is given by
$d \sigma_i/dp_t$.
Then the jet spectrum will be given by
\begin{equation}
  \label{eq:inclusive-microjet-spectrum}
  \frac{d\sigma_\text{jet}}{dp_t} 
  = \sum_i \int dp_{t}' dz 
  \frac{d\sigma_i}{dp_t'} 
  f^{\text{incl}}_{\text{jet}/i}(z,t)
  \delta(p_t - z p_t')
  = \sum_i \int_{p_t} \frac{dz}{z}
  \left.\frac{d\sigma_i}{dp_t'}\right|_{p_t' = p_t/z} 
  f^{\text{incl}}_{\text{jet}/i}(z,t)\,,
\end{equation}
where $f_{{\rm jet}/i} \equiv\sum_j f_{j/i}$.
If we assume that the partonic spectrum is dominated by a single
flavour $i$ and its $p_t$ dependence locally is $d\sigma_i/dp_t \sim
p_t^{-n}$, then one obtains the following multiplicative relation
between the microjet and partonic spectra,
\begin{equation}
  \label{eq:inclusive-microjet-moments}
  \frac{d\sigma_\text{jet}}{dp_t} 
  \simeq \frac{d\sigma_i}{dp_t} \int_0^1 dz \, z^{n-1}\,
  f^{\text{incl}}_{\text{jet}/i}(z,t)\,.
\end{equation}
We will use the shorthand $\langle
z^{n-1}\rangle_i^\text{incl}$ for the $z$ integral here.
The all-order results are trivial to obtain analytically from the
exponentiation of the matrix of moments of LO DGLAP splitting
functions. 
The order-by-order expansion is likewise trivial.
We accordingly refer the reader to Eqs.~(4.128) and (4.129) of
Ref.~\cite{Ellis:1991qj}.

\begin{figure}[t]
  \centering
  \includegraphics[width=0.50\textwidth]{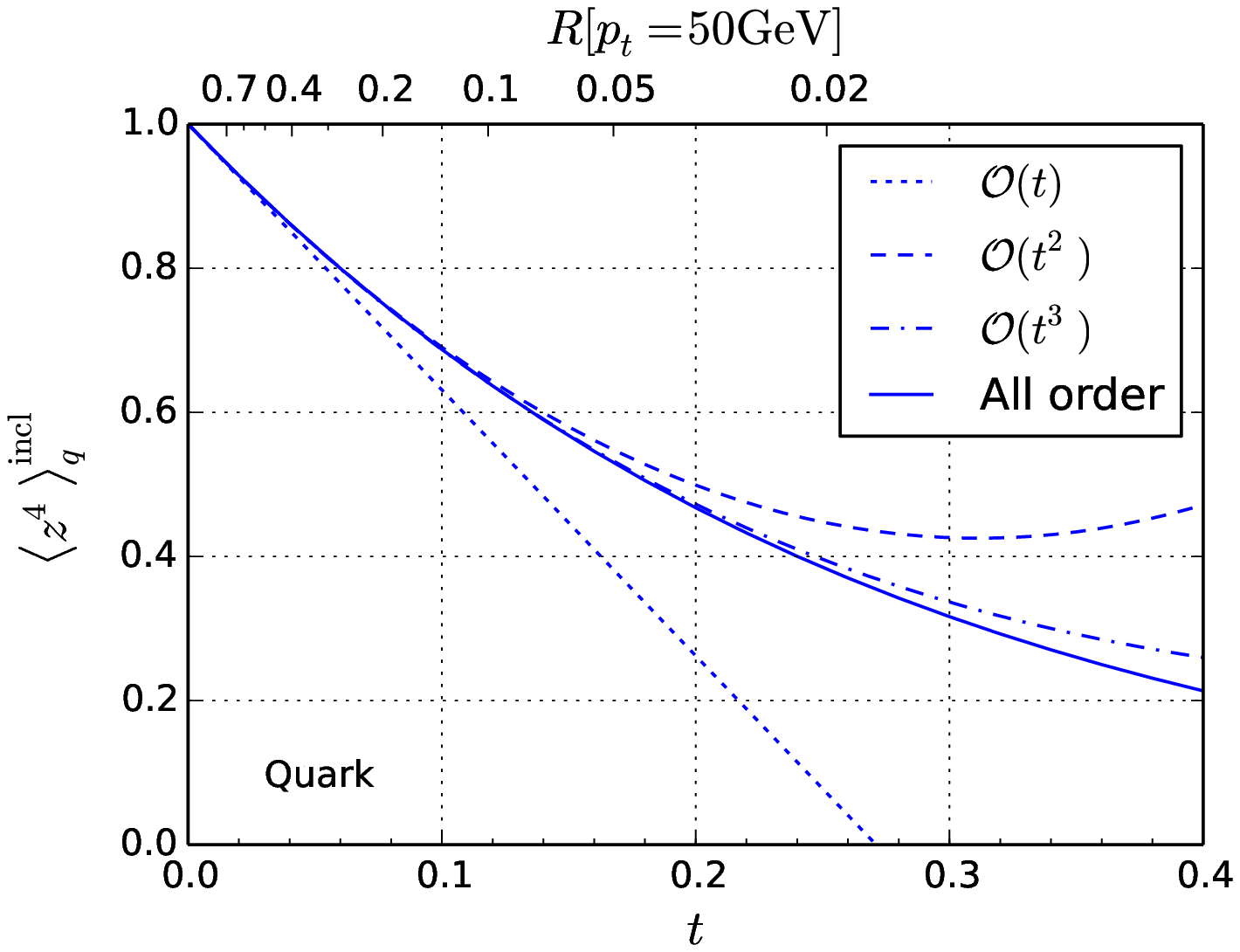}%
  \includegraphics[width=0.50\textwidth]{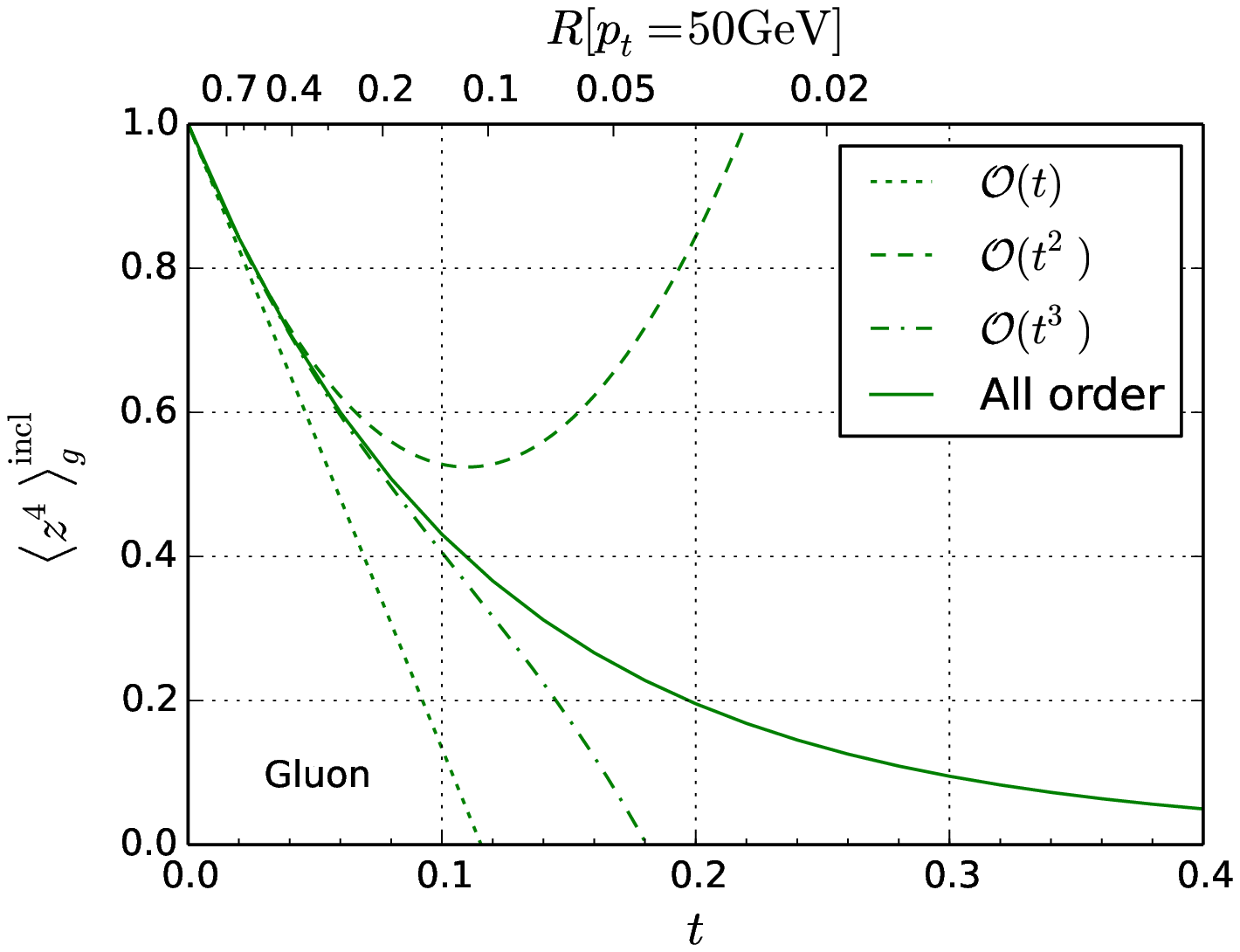}
  \caption{The result for $\langle z^4\rangle^\text{incl}$ at all
    orders as a function of $t$ (lower axis), together with the first
    3 orders of its expansion in $t$, shown for initiating quarks
    (left) and gluons (right).
    The upper axis gives the correspond $R$ values for a jet with
    $p_t$ of order $50\GeV$.  
    The factor $\langle z^4\rangle^\text{incl}$, multiplied by a hard
    inclusive parton spectrum that goes as $p_t^{-5}$, gives the
    corresponding microjet spectrum.  
    \label{fig:incl-pt5}
  }
\end{figure}

At the LHC, typical $n$ values range from about $4$ at low $p_t$ to
$7$ or even higher at high $p_t$.
The resummed $(n\!-\!1)^\text{th}$ moment of the inclusive microjet
fragmentation function is shown for quarks and gluons in
Fig.~\ref{fig:incl-pt5}, for $n=5$, together with the first few orders
of the perturbative expansion in $t$.
In this, and most of the other plots that follow, the lower $x$-axis
shows the value of $t$, while the upper axis shows the corresponding
value of $R$ for the case of a parton with $p_t = 50\GeV$.

The first observation that one makes from Fig.~\ref{fig:incl-pt5} is
that small-$R$ effects can be substantial. 
For quark-induced jets with $R$ in the range $0.4$ to $0.2$, they
reduce the inclusive-jet spectrum by $15-25\%$.
For gluon-induced jets the corresponding reductions are $30-50\%$.
These are substantial reductions in the cross-section, and help provide a
motivation for wanting to understand small-$R$ effects.

One question one can ask is about the convergence of the perturbative
series.
For quark-induced jets the $\order{t}$ (i.e.\ NLO) result is accurate
to within a couple of percent for $R=0.4$, while at $R=0.2$ one sees
$5\%$ differences relative to the all-order result.
For gluon-induced jets, the $\order{t}$ result is off by about $10\%$ 
for $R=0.4$, becomes inadequate around $R=0.3$ and
pathological (negative) near $R=0.1$.
Including the $\order{t^2}$ corrections (i.e.\ NNLO) brings agreement
with all-orders to within a couple of percent for quark-initiated jets
down to $R=0.1$; for gluon-initiated jets $\order{t^2}$ is adequate at
$R=0.4$, but starts to deviate noticeably from the all-orders results
below $R=0.3$.

Note that an expansion in $t$ is not directly equivalent to an
expansion in $\alpha_s$, because the variable $t$ already resums the
running-coupling contributions.
As we will discuss at more length in a companion paper
\cite{smallR-pheno}, an expansion in $\alpha_s$ appears more
convergent. 
However, in light of the pattern of corrections as a power series in
the natural evolution variable $t$, one wonders whether this
apparently better convergence in $\alpha_s$ is to be trusted.

%----------------------------------------------------------------------
\subsection{Hardest microjet observables}

As well as the inclusive jet spectrum, it is common to ask questions
about the hardest jet in an event, i.e.\ the jet with the largest
transverse momentum.
The hardest jet $p_t$ observable is relevant also whenever a jet veto
is applied, since a veto is equivalent to a requirement that the
hardest jet's $p_t$ be below some threshold.
Given a parton with transverse momentum $p_t$, we define
$f^\text{hardest}(z)$ to be the probability that the hardest resulting
microjet carries a momentum $z p_t$.\footnote{We should really write
  $f^\text{hardest}(z,t)$, but in most cases drop the $t$ argument for
  compactness.}
Now instead of a momentum sum rule, we have a probability sum rule,
\begin{equation}
  \label{eq:fhardest-prob}
  \int_0^1 dz \, f^\text{hardest}(z) = 1
\end{equation}
In general, quantities involving the hardest microjet are more
complicated than the inclusive quantities considered in
section~\ref{sec:incl-microjet}.
The reason is that for an ensemble of microjets arising from multiple
nested splittings, one has to consider all of the microjets together
in order to determine which is the hardest.
We have been able to carry out analytical calculations for
configurations with up to three partons (order $t^2$), but have
resorted to Monte Carlo methods to evaluate higher orders in $t$ and
all-order results.

The all-order distribution of $f^\text{hardest}(z)$ is shown as dashed lines
in Fig.~\ref{fig:fragfct}, for the same four $t$ values as the
inclusive microjet fragmentation function.
The inclusive and hardest microjet spectra are
identical for $z>\frac12$, since it is impossible to have more than one
microjet with $z > \frac12$.
For small values of $t$, the hardest microjet spectrum shows a sharp
transition at $z=\frac12$, because
below $z=\frac12$ it can be non-zero only starting at order
$t^2$. 
There is another transition at $z = \frac13$, below which there are
contributions only from order $t^3$ onwards.
As $t$ increases, these transition points smoothen out significantly. 
The main difference between quark and gluon initiated jets is that the
effects are more marked for the latter, as one would expect from
the larger colour factor.

Next we consider various average properties of the hardest microjet.
As well as numerical resummed results, we will also provide the
coefficients of the first few orders of the power series in $t$, which
we write as
\begin{equation}
  \label{eq:pert-exp}
  \mathcal{O}(t)=\sum_n\frac{t^n}{n!}\, c_n[\mathcal{O}]\,,
\end{equation}
for a general observable $\mathcal{O}$.
While we have analytical results for $c_2$, for the sake of
conciseness we will just quote numerical values in this section,
separately for each colour factor. 
The corresponding full analytical expressions are given in
appendix~\ref{sec:appendix-analytic}.

%......................................................................
\subsubsection[Hardest microjet ${\langle \Delta z \rangle}$]{Hardest microjet $\boldsymbol{\langle \Delta z \rangle}$}
\label{sec:delta-p_t}

A typical context in which the average fractional energy loss from a jet,
$\langle \Delta z \rangle$, is relevant is in the study of the
difference in $p_t$ between a $Z$-boson and the leading jet in $Z+$jet
events.
This kind of quantity is used for jet
calibration~\cite{Aad:2014bia,Chatrchyan:2011ds}.
It is also relevant for example in jet--photon balance studies
in heavy-ion collisions~\cite{ATLAS:2012cna,Chatrchyan:2012gt}.

The average fractional transverse-momentum difference between the
hardest microjet and the initial parton is given by
\begin{equation} \label{eq:def-deltapt}
  \langle \Delta z \rangle ^\text{hardest}
  \equiv \int_0^1\, dz\,
  f^\text{hardest}(z) (z-1) \,.
\end{equation}
For an initial quark we find
\begin{multline} \label{eq:deltapt-Q}
  \langle \Delta z \rangle^\text{hardest}_q 
  = C_F t \left(
      \frac{3}{8}-2\ln 2
 \right) + 
 \\
  +\frac{t^2}{2}\big( -0.467188 C_A C_F 
      + 1.62588 C_F^2
      - 0.0710364 C_F
      n_f T_R\big)\, +
  \\
  % values for seed=1,...,10
  + \frac{t^3}{6}\big( - 2.33574(2) C_F^3 
      + 0.67962(2) C_A^2 C_F 
      + 0.11881(2) C_A C_F^2 
      + 0.416131(6) C_A C_F n_f T_R -
      \\
      - 0.204121(5) C_F^2 n_f T_R 
      + 0.0473591(7) C_F n_f^2 T_R^2 \big) 
  + \mathcal{O}(t^4)\,,
\end{multline}
while the case of an initiating gluon yields
\begin{multline}\label{eq:deltapt-G}
  \langle \Delta z \rangle^\text{hardest}_g 
  = t\left[
      -\frac{7}{48} n_f T_R 
      + C_A\left(\frac{43}{96} - 2\ln 2\right)
  \right] +
  \\
  +\frac{t^2}{2} \big( 0.962984 C_A^2 
      + 0.778515 C_A n_f T_R 
      - 0.50674 C_F n_f T_R 
      + 0.0972222 n_f^2 T_R^2 \big)\, +
  \\
  % values with seed=1,...,10
  +\frac{t^3}{6}\big( - 1.11718(2) C_A^3 
      - 1.557542(7) C_A^2 n_f T_R 
      + 0.375492(7) C_A C_F n_f T_R
      + 0.75869(1) C_F^2 n_f T_R -
      \\
      - 0.635406(3) C_A n_f^2 T_R^2 
      + 0.305404(3) C_F n_f^2 T_R^2 
      - 0.0648152(4) n_f^3 T_R^3 \big)
  + \mathcal{O}(t^4)\,.
\end{multline}

\begin{figure}[t]
  \centering
  \includegraphics[width=0.50\textwidth]{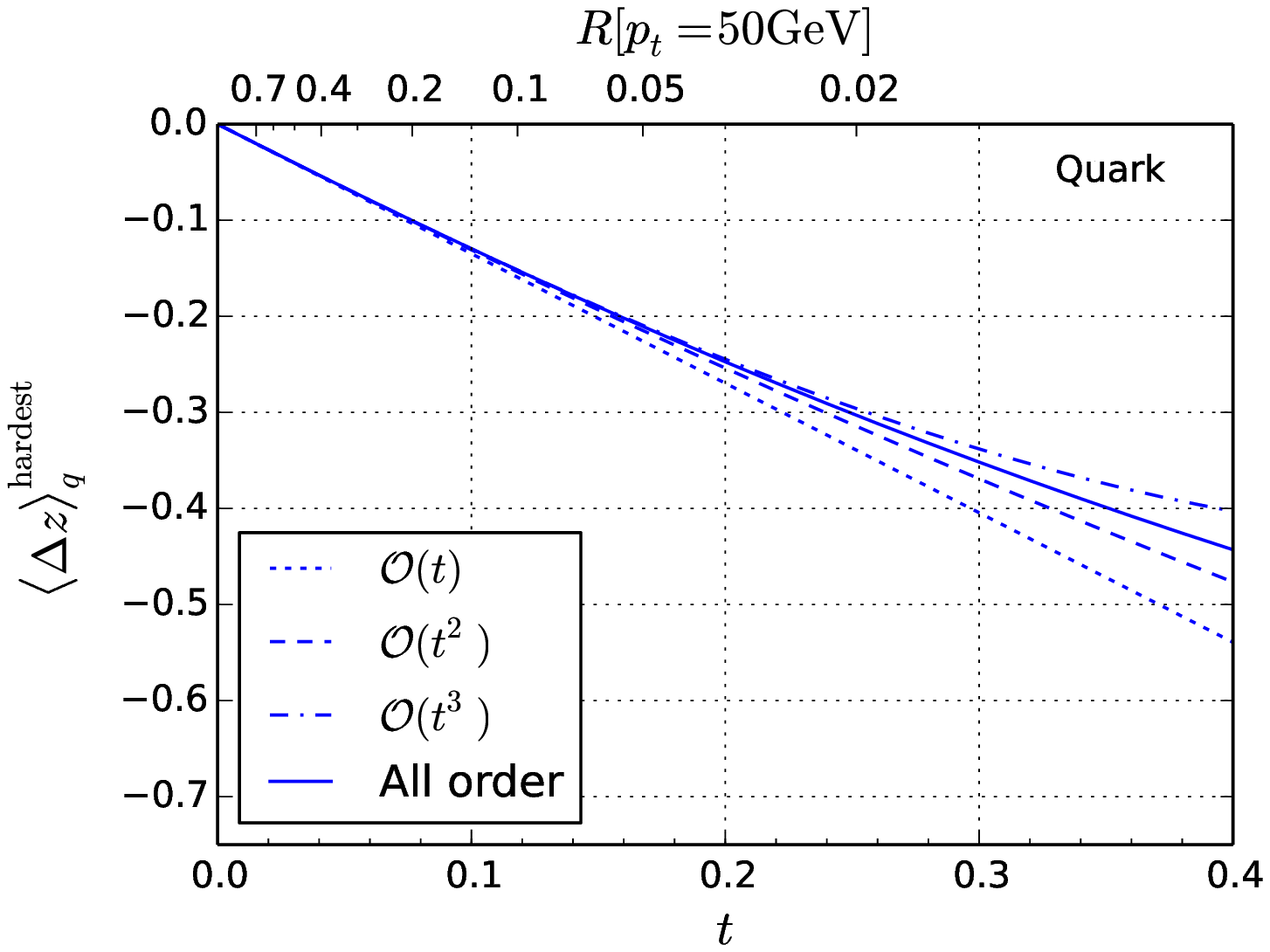}%
  \includegraphics[width=0.50\textwidth]{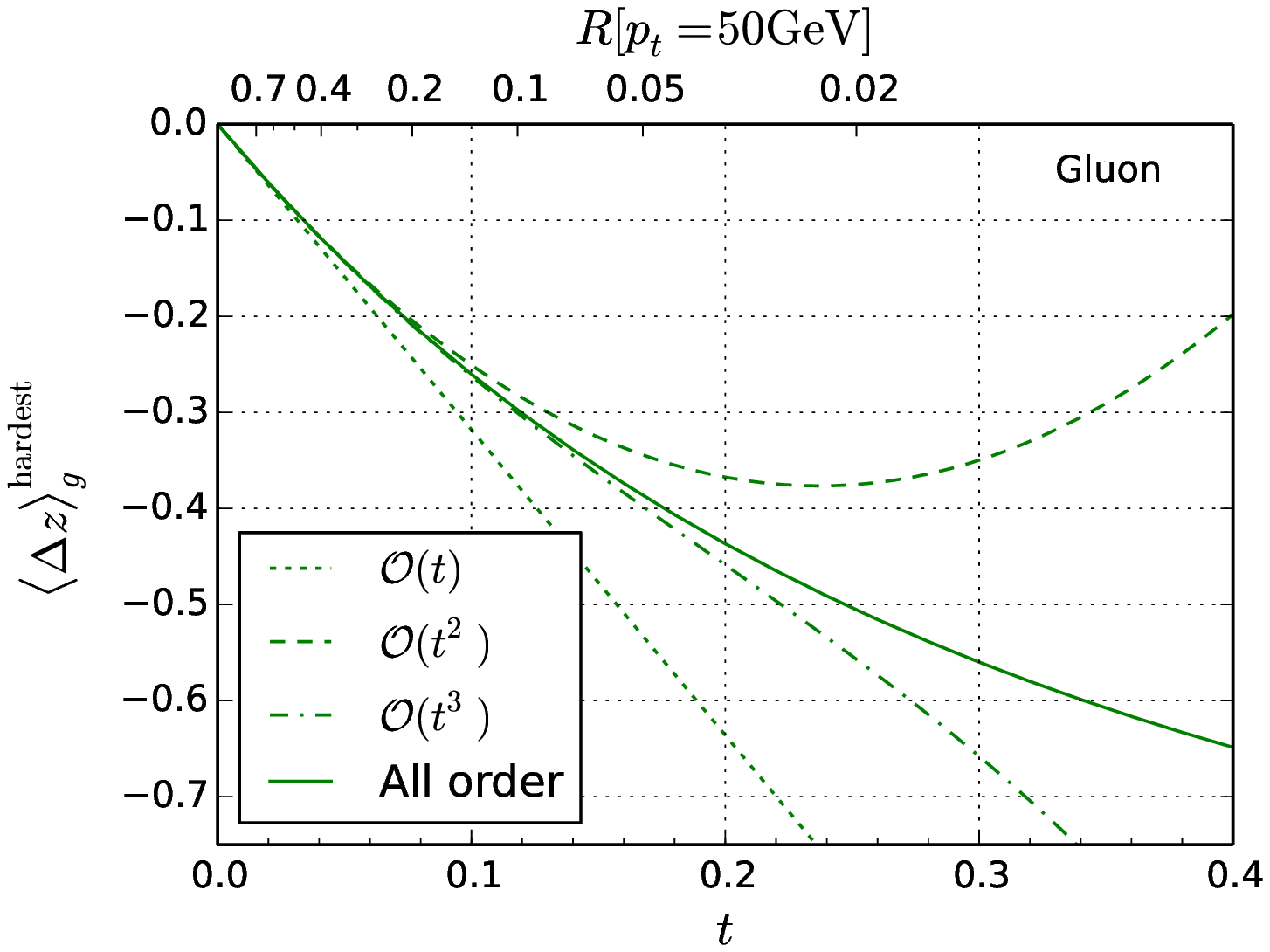}%
  \caption{Average hardest microjet $\Delta z$, shown as a function of
    $t$ for quark-induced (left) and gluon-induced (right) jets.
    Resummed results are represented as solid
    lines. 
    The first three orders in $t$ are represented as
    dotted, dashed and dash-dotted lines respectively.
    \label{fig:deltapt}
  }
\end{figure}

In figure~\ref{fig:deltapt} we can see the all-order results for the
hardest microjet $\Delta z$, along with the fixed order expansion
expressed in equations~(\ref{eq:deltapt-Q}) and (\ref{eq:deltapt-G})
truncated at the first, second and third powers in $t$.
For quark-induced jets, the fractional energy loss is in the range
$5-10\%$ for $R=0.4-0.2$, while for gluon-induced jets it is in the
range $10-20\%$.

One feature of Fig.~\ref{fig:deltapt} is that one immediately notices
a significantly better convergence than in Fig.~\ref{fig:incl-pt5}.
This is because the $z^4$ weighting in Fig.~\ref{fig:incl-pt5}
amplifies the impact of higher orders.
On the other hand jet momenta tend to be measured with much higher
accuracy ($\sim 1\%$~\cite{Aad:2014bia,Chatrchyan:2011ds}) than
steeply falling jet spectra, so one targets higher relative accuracy
for $\langle \Delta z \rangle^\text{hardest}$.
Quite high (percent-level) accuracy for the phenomenologically
relevant range of $t$ is obtained even at order $t$ in the case of
quark jets.
However, for gluon jets $\order{t}$ is probably inadequate at $R=0.3$
and below.
Going to order $t^2$ is probably sufficient for gluon jets down to
$R=0.15$.
As noted in the case of the inclusive jet spectrum an expansion in $t$
does not directly correspond to an expansion in $\alpha_s$, so further
cross-checks on the validity of fixed-order calculations would be
needed on a case-by-case basis.

%......................................................................
\subsubsection[Jet-veto resummations and ${\langle \ln z
  \rangle}$]{Jet-veto resummations and $\boldsymbol{\langle \ln z
    \rangle}$} 
\label{sec:log-p_t}

Jet veto resummations are one of the contexts in which the potential
need for all-order small-$R$ corrections has been
raised~\cite{Tackmann:2012bt}.
In this context, as we shall derive below, the relevant quantity is
$\langle \ln z \rangle^\text{hardest}$.

Let us first recall the core structure of a jet-veto resummation
for finite-$R$ jets.
As we shall see, when considering logarithms of the jet veto scale, it
will be sufficient for our purposes to work at leading (double)
logarithmic accuracy.
This helps eliminate numerous complications
such as those related to parton distribution functions.
Assuming a process with a hard scale $Q$ and two incoming partons with
colour factor $C$ ($C_F$ for quarks, $C_A$ for gluons), the
probability of there being no gluons emitted above a scale $p_t$ is
given by
\begin{equation}
\label{eq:jet-veto-probability-partons-start}
  P(\text{no primary-parton veto}) = \sum_{n=0}^\infty \frac{1}{n!}
  \prod_{i=1}^n \left[\int^Q \frac{dk_{ti}}{k_{ti}} 
    \asbar(k_{ti})\,  2\ln \frac{Q}{k_{ti}}
    \left(-1 + \Theta(p_t - k_{ti}) \right) \right]\,,
\end{equation}
where $\asbar(k_t) \equiv 2 \as(k_t) C/\pi$.
In the factor $\left(-1 + \Theta(k_{ti} - p_t) \right)$, the
$\Theta$-function corresponds to the veto on partons above a scale
$p_t$, while the term $-1$ accounts for virtual corrections.
The factor $2\ln \frac{Q}{k_t}$ corresponds to the kinematically
allowed range of rapidities for a gluon with transverse momentum
$k_t$ (in a leading logarithmic approximation for the $p_t$ veto
resummation).
It is straightforward to see that
Eq.~(\ref{eq:jet-veto-probability-partons-start}) corresponds to an
exponential, 
\begin{equation}
\label{eq:jet-veto-probability-partons-exp}
  P(\text{no primary-parton veto}) = 
  \exp
  \left[-\int^Q_{p_t} \frac{dk_{t}}{k_{t}} 
    \asbar(k_{t})\,  2\ln \frac{Q}{k_t}\right]\,.
\end{equation}
Defining $L \equiv \ln Q/p_t$, $\ln P$ contains ``leading (double)
logarithmic'' terms $\as^n L^{n+1}$.
The jet veto efficiency in Higgs and Drell-Yan production is
currently known to
NNLL accuracy in this language, i.e.\ $\as^n
L^{n-1}$~\cite{Banfi:2012yh,Banfi:2012jm,Becher:2012qa,Becher:2013xia,Stewart:2013faa}. 
The papers by the Becher et al.\ and the Stewart et al.\ groups
include subsets of terms beyond NNLL, while heavy-quark effects in the
$ggH$ interaction have been discussed in Ref.~\cite{Banfi:2013eda}.

To include small-$R$ corrections, one needs to modify
Eq.~(\ref{eq:jet-veto-probability-partons-start}) to account for the
fact that each of the partons $i=1\ldots n$ will fragment into
multiple microjets, and for each of those partons the veto now applies
to the resulting hardest microjet.
Thus for each emission in
Eq.~(\ref{eq:jet-veto-probability-partons-start}), we integrate over
the probability distribution for the momentum fraction of the hardest
resulting microjet,
\begin{multline}
\label{eq:jet-veto-probability-microjet-start}
  P(\text{no microjet veto}) = \sum_{n=0}^\infty \frac{1}{n!}
  \prod_{i=1}^n \left[\int^Q \frac{dk_{ti}}{k_{ti}} 
    \asbar(k_{ti})\,  2\ln \frac{Q}{k_{ti}}
    \,\times
    \right.
    \\
    \left.
    \times \int_0^1 dz_i f^\text{hardest}(z_i,t(R,k_{ti}))
    \Bigg(-1 + \Theta(p_t - z_i k_{ti}) \Bigg) \right]\,,
\end{multline}
where we have made the $t(R,k_{ti})$ argument in
$f^\text{hardest}(z_i,t(R,k_{ti}))$ explicit, because of the
importance of making the right scale choice for the definition of $t$
(cf.\ also Eq.~(\ref{eq:t})).
Given Eq.~(\ref{eq:fhardest-prob}), it is immaterial to the result
whether the integration over $z$ takes place outside the large round
brackets or inside, applied just to the $\Theta$-function.
As before, we now write the result as an exponential,
\begin{multline}
  \label{eq:jet-veto-probability-microjet-exp}
  P(\text{no microjet veto}) = 
  \exp
  \left[- 
    \int^Q \frac{dk_{t}}{k_{t}} 
    \asbar(k_{t})\,  2\ln \frac{Q}{k_t}
    \times \right.
    \\
    \left. \times
    \int_0^1 dz f^\text{hardest}(z,t(R,k_{t}))
    \big(
    \Theta(k_{t} - p_t)
    + 
    \Theta(z k_{t} - p_t)
    - 
    \Theta(k_{t} - p_t)
     \big)
  \right].
\end{multline}
To help us perform the calculation, we have added and subtracted a term
$\Theta(k_t - p_t)$.
Then one considers the first $\Theta$ function separately from the
second and third.
%We have separated the $\Theta$ functions into multiple pieces: f
%
For the
first one, the $z$ integration can be performed trivially and one obtains
the primary-parton result,
Eq.~(\ref{eq:jet-veto-probability-partons-exp}). 
For the remaining pair of $\Theta$-functions, we first evaluate the
$k_t$ integral: since $k_t$ is being evaluated over a limited range,
we can replace $\asbar(k_t)$ with $\asbar(p_t)$, and similarly in the
$\ln Q/k_t$ and $t(R,k_t)$ factors.
The terms that we neglect as a result of this are suppressed by
one logarithm of $\ln Q/p_t$.
We therefore have the following small-$R$ correction to the jet veto
efficiency
\begin{equation}
  \label{eq:jet-veto-probability-microjet-processed}
  \mathcal{U} \equiv \frac{P(\text{no microjet veto})}{P(\text{no primary-parton veto})} = 
  \exp
  \left[-
      2 \asbar(p_t) \ln \frac{Q}{p_t}
      \int_0^1 dz f^\text{hardest}(z,t(R,p_{t})) \ln z
    \right].
\end{equation}
This $R$-dependent correction generates a series of terms $\as^{m+n}(Q)
L^m \ln^n R^2$, while we have neglected terms suppressed by one or
more powers of either $L= \ln Q/p_t$ or $\ln R^2$. 

Eq.~(\ref{eq:jet-veto-probability-microjet-processed}) shows that the
key quantity for the small-$R$ part of the resummation is the first
logarithmic moment of $f^\text{hardest}(z)$
\begin{equation}\label{eq:def-logpt}
  \langle \ln z \rangle^\text{hardest} \equiv \int_0^1\, dz\,
  f^\text{hardest}(z) \ln z \,.
\end{equation}
It is the logarithmic moment of the microjet spectrum
from initial gluons that is relevant here, since the $i=1\ldots n$ partons in
Eq.~(\ref{eq:jet-veto-probability-microjet-start}) are all gluons
(regardless of whether the jet veto is applied to a $q\bar q$ or
gluon-fusion process).
The first three orders of its expansion in $t$ are
\begin{multline}\label{eq:logpt-G}
  \langle \ln z \rangle_g^\text{hardest} = t\bigg[
      \frac{1}{72} C_A \left(131 - 12 \pi ^2 - 132\ln 2\right)
      + \frac{1}{36} n_f T_R (-23+24 \ln 2)
  \bigg] + 
  \\
  + \frac{t^2}{2} \big(  0.206672 C_A^2 
      + 0.771751 C_A n_f T_R 
      - 0.739641 C_F n_f T_R 
      + 0.117861 n_f^2 T_R^2 \big)\, +
  \\
  % values with seed = 1,..10
  + \frac{t^3}{6} \big(-0.20228(4) C_A^3 
      - 0.53612(2) C_A^2 n_f T_R 
      - 0.062679(8) C_A C_F n_f T_R 
      + 0.54199(2) C_F^2 n_f T_R\, -
      \\
      - 0.577215(3) C_A n_f^2 T_R^2 
      + 0.431055(4) C_F n_f^2 T_R^2 
      - 0.0785743(5) n_f^3 T_R^3 \big)
   + \mathcal{O}(t^4) \,.
\end{multline}
The order $t$ term of Eq.~(\ref{eq:logpt-G}), when incorporated into
Eq.~(\ref{eq:jet-veto-probability-microjet-processed}), gives an $\as^2
L \ln R^2$ term whose coefficient agrees with that given in
Refs.~\cite{Banfi:2012jm,Becher:2013xia,Stewart:2013faa}. 
Full analytical results for the $\order{t^2}$ term are to be found in
Appendix~\ref{sec:app-ln-pt}, as are analytical and numerical results
for the corresponding logarithmic moment for quark fragmentation
(potentially of interest for vetoes on initial state $g\to q\bar q$
splittings).

\begin{figure}[t]
  \centering
  \includegraphics[width=0.50\textwidth]{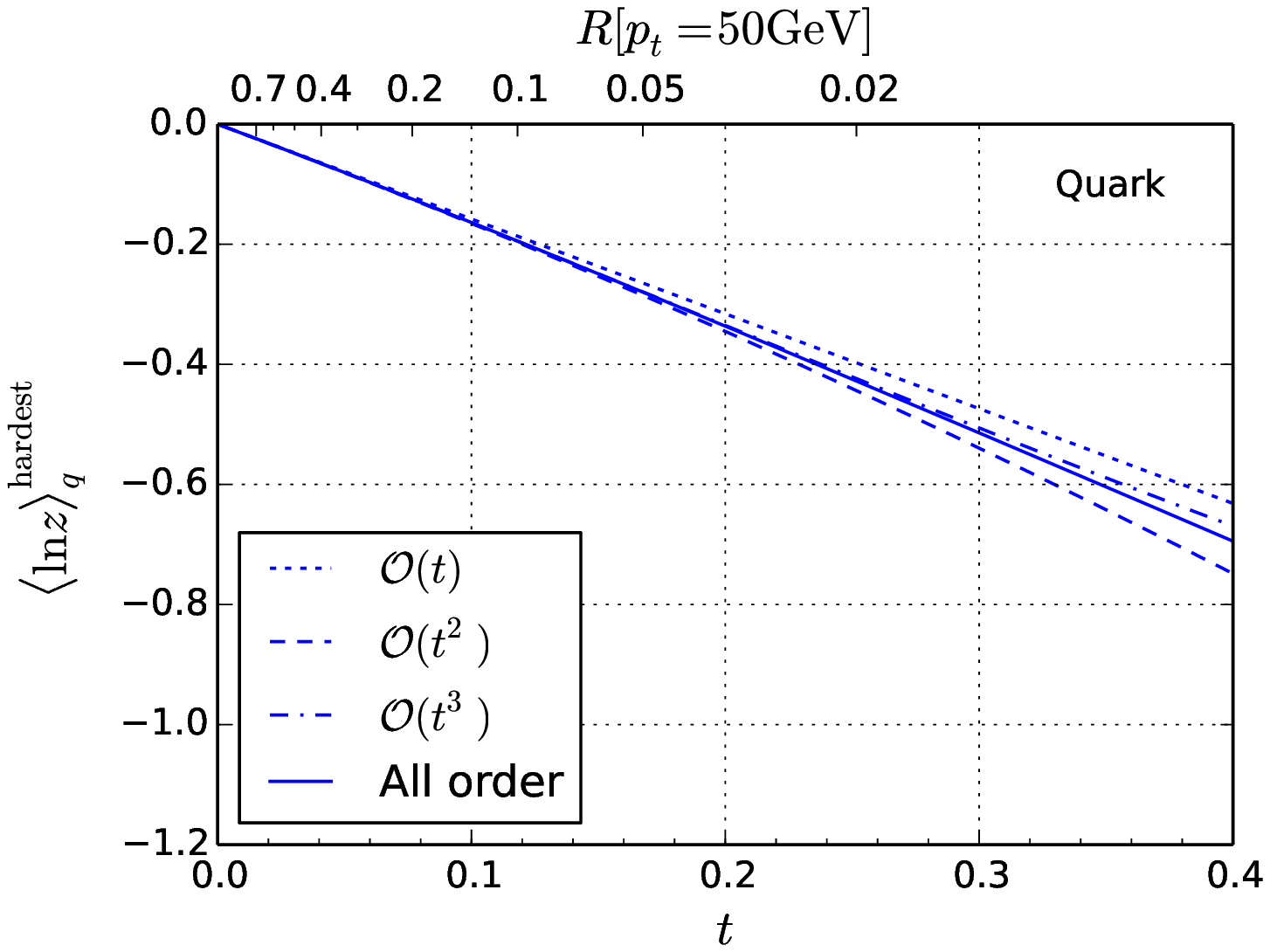}%
  \includegraphics[width=0.50\textwidth]{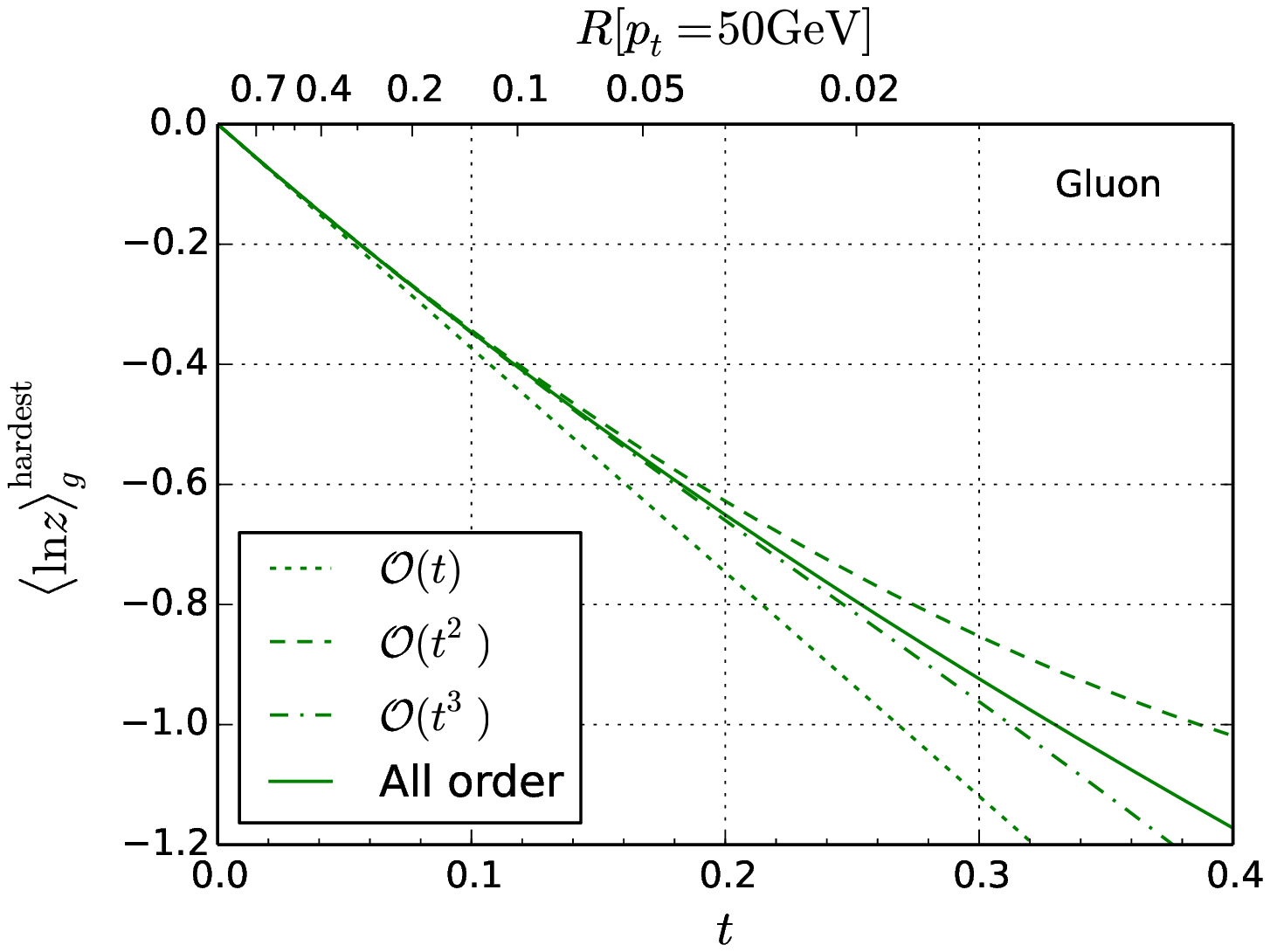}%
  \caption{Average of the hardest microjet $\ln z$, as a function of
    t, shown separately for quark-induced (left) and gluon-induced (right) jets.
    The  resummed results are
    represented as solid lines. The 
    first three orders in $t$ are represented as dotted, dashed and
    dash-dotted lines respectively.
    \label{fig:logpt}
  }
\end{figure}

\begin{table}
  \centering
  \begin{tabular}{cccccc}
    \toprule
    $\mathcal{O}$ & $c_1$ & $c_2$ & $c_3$ &
    $c_4^\textrm{fit}$ & $c_5^\textrm{fit}$
    \\
    \midrule
    $\langle\ln z\rangle_q^\text{hardest}$ & -1.58  & -1.46 & 7.43 & -34.07 & 105.62 \\[4pt]
    $\langle\ln z\rangle_g^\text{hardest}$ & -3.73 & 5.92 & -24.20 & 133.55 & -478.55\\[2pt]
    \bottomrule
  \end{tabular}
  \caption{Results of a fit to the all-order determination of
    $\langle\ln z\rangle^\text{hardest}$; the fit function is of the
    form $c_1 t + c_2 \frac{t^2}{2!} 
    + c_3 \frac{t^3}{3!} 
    + c_4^\textrm{fit} \frac{t^4}{4!} 
    + c_5^\textrm{fit} \frac{t^5}{5!} $\,
    where  the $c_1$, $c_2$ and $c_3$ coefficients are fixed to the values given in
    Eqs.~(\ref{eq:logpt-Q}) and (\ref{eq:logpt-G}), while
    $c_4^\textrm{fit}$ and $c_5^\textrm{fit}$ have been fitted.
    The fit was performed in the range $0<t<0.4$ and has an accuracy
    of significantly better than $1\%$ for $\langle \ln
    z\rangle^\text{hardest}$.  
    The fit values for the fourth and fifth order coefficients are not
    to be taken as robust determinations of those coefficients, but
    simply as values whose use in the truncated sum gives good
    agreement with the all-order result.
  }
  \label{tab:jetvheto-fit}
\end{table}

The all-order result for $\langle \ln z \rangle_g^\text{hardest}$ (and
quark counterpart) is shown as a function of $t$ in
Fig.~\ref{fig:logpt}, together with its expansion in $t$.
The small-$R$ effects in the gluon-induced case are in the range
$15\%$ to a little over $25\%$ for $R=0.2-0.4$, i.e.\ slightly larger than for
$\langle \Delta z \rangle_g^\text{hardest}$.
One notes the remarkably good convergence: for example at $t=0.1$, for
$\langle \Delta z \rangle_g^\text{hardest}$ the difference between the
$\order{t}$ and all-order results was roughly $25\%$; in contrast for
$\langle \ln z \rangle_g^\text{hardest}$ the difference is $7\%$.
It is not clear to us if there is a fundamental reason why this should
be the case.

For practical use it is useful to have a parametrisation of the
all-order result. 
%915
Given the good convergence of the series, this can be obtained in the
range $t<0.4$ simply by fitting additional $c_4$ and $c_5$
coefficients to the all-order curve. 
The results of the fit are given in table~\ref{tab:jetvheto-fit} and
they allow one to reproduce the all-order result to an accuracy of
better than $1\%$.
The actual values of the coefficients themselves are, however, not
guaranteed to be accurate since they may in part be absorbing
contributions from yet higher order terms.
A study of the phenomenological impact of the small-$R$ resummation
will be given elsewhere~\cite{smallR-pheno}.

We note that a numerical calculation for the $\as^3 L \ln^2 R^2$ term
in the case of the jet veto for $gg\to H$ production was given in
Ref.~\cite{Alioli:2013hba}.
At first sight it appears to disagree with our analytical result and
after consultation with the authors they identified an issue in their
treatment of $R$-dependent running coupling-related terms.
The detailed comparison and discussion is to be found in
Appendix~\ref{sec:alioli-walsh}.
In Appendix~\ref{sec:fixed-order-cross-checks} we present a
cross-check of our calculational approach specifically for the
second-order contribution to $f_{\text{hardest}}(z)$ for quark-induced
jets, as obtained through a comparison to the Event~2 NLO
program~\cite{Catani:1996vz}.

%......................................................................
\subsubsection{Jet flavour}
\label{sec:jet-flav}

The flavour of jets is a subject that is conceptually interesting and
obtaining a better handle on jet flavour is potentially also of
considerable practical use.\footnote{Furthermore, jet flavour has seen
  extensive discussion in the literature, so much so that, for
  example, Ref.~\cite{Banfi:2006hf} was able to identify over $350$
  articles with the terms ``quark jet'' or ``gluon jet'' in their
  title.}
As discussed at length in Ref.~\cite{Banfi:2006hf}, the definition of
jet flavour is a subtle question. 
However, in the leading-logarithmic collinear limit in which we work
here, those subtleties disappear, essentially because they are related
to soft radiation. 

The question that we ask in this subsection is the following: given a
quark (gluon) parton, how likely is it that the resulting hardest
microjet will have the flavour of a gluon (quark).
This is relevant, for example, when considering the performance of
quark/gluon tagging algorithms, whether
theoretically~\cite{Gallicchio:2011xq,Gallicchio:2012ez,Larkoski:2013eya,Larkoski:2014pca}
or experimentally~\cite{Aad:2014gea,CMS:2013kfa}, since one is
often assuming that the flavour of the selected (hardest) jet is
identical to that of the underlying hard scattering.

\begin{figure}[tp]
  \centering
  \includegraphics[width=0.5\textwidth]{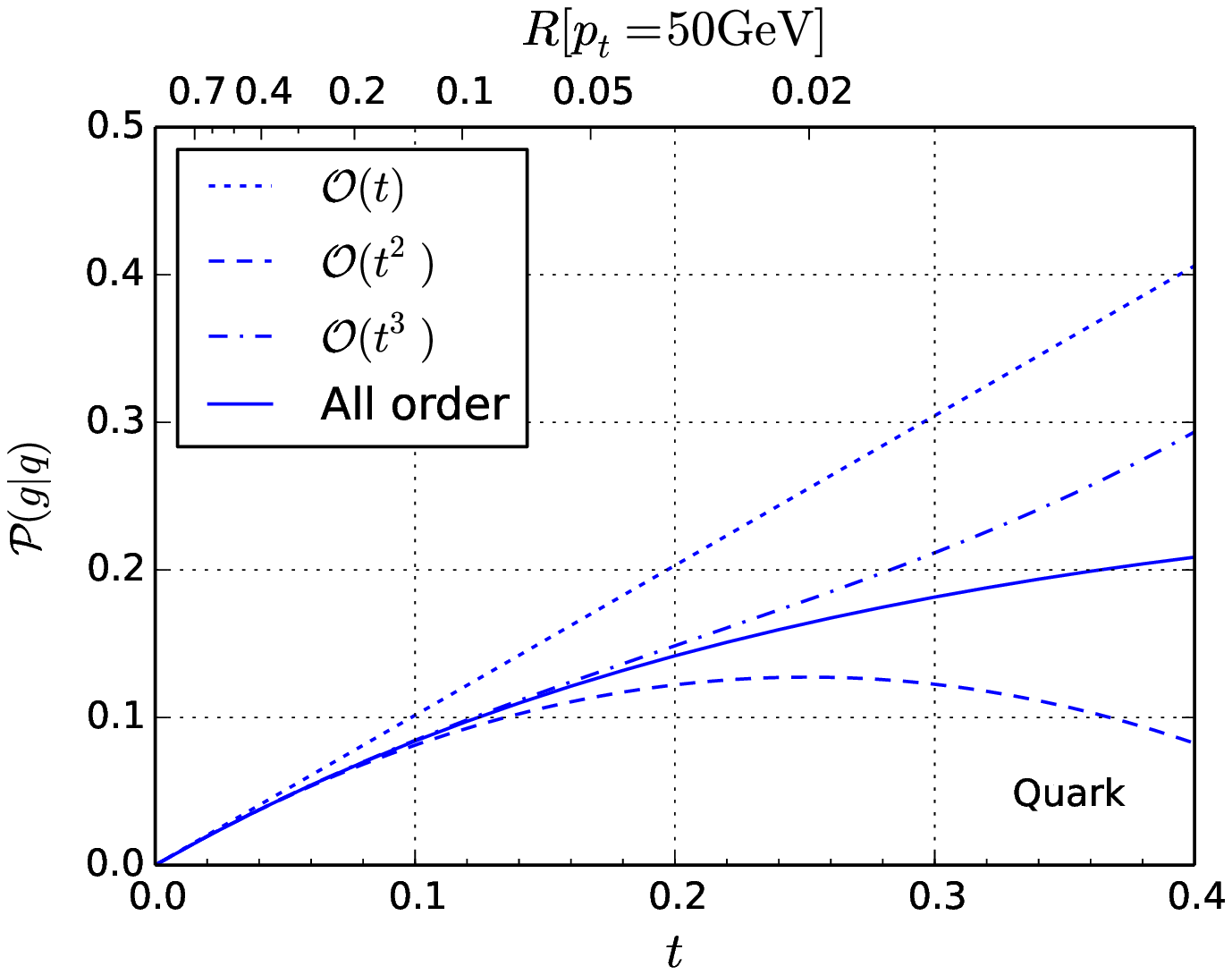}%
  \includegraphics[width=0.5\textwidth]{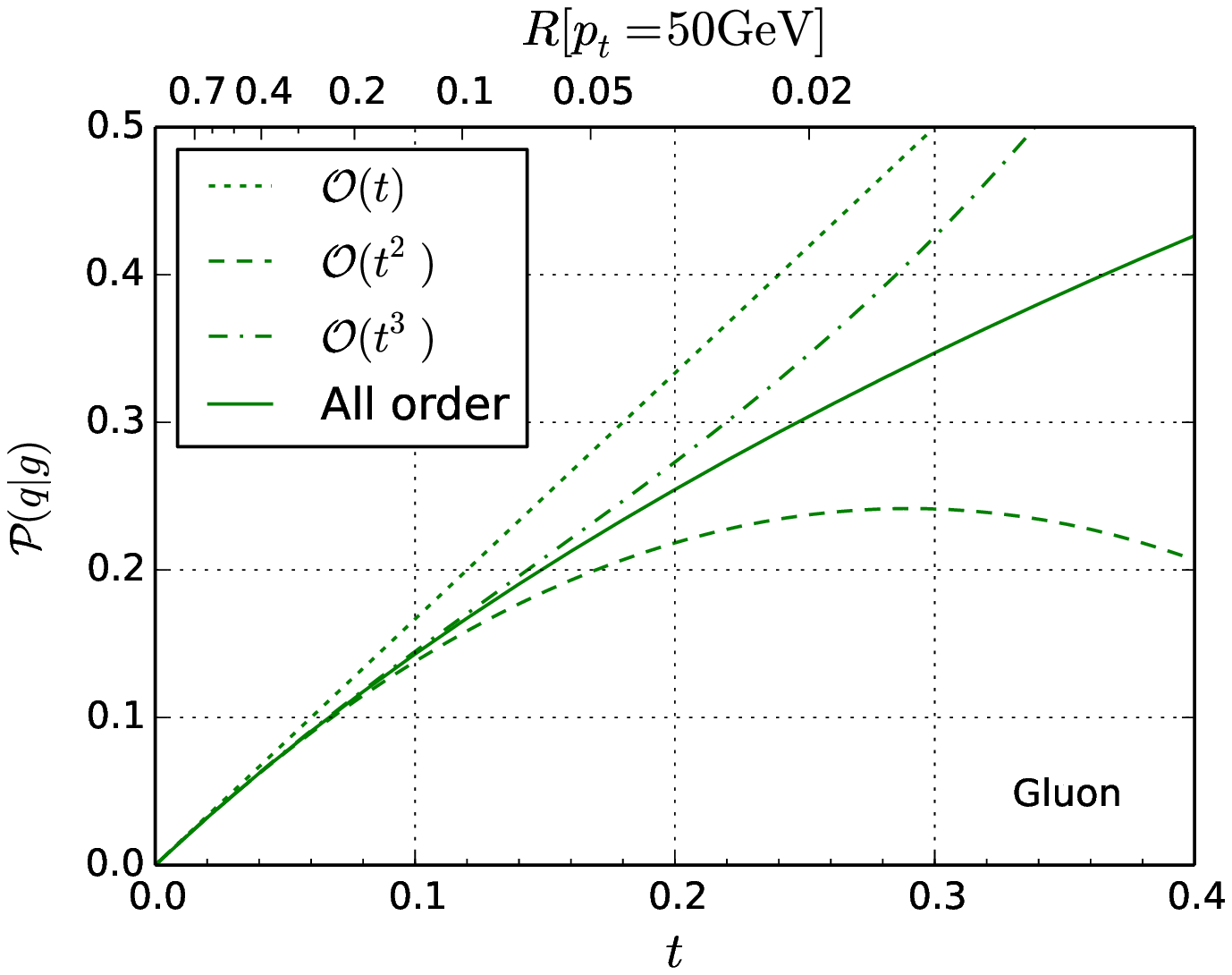}
  \caption{Hardest microjet flavour change probability as a function
    of $t$. 
    The left-hand plot shows the probability for a quark-parton to
    emerge as a gluon microjet, while the right-hand plot shows the
    opposite flavour-changing probability.
    Resummed results are shown as solid lines, while the first
    three orders in $t$ are represented as dotted, dashed and
    dash-dotted lines respectively.
    \label{fig:flavchange}
  }
\end{figure}

To answer this question, we extend $f^\text{hardest}(z)$ to have
flavour indices: $f_{a/b}^\text{hardest}(z)$ is the differential
distribution in $z$ of hardest microjets of flavour $a$ given an
initiating parton of flavour $b$.
The overall probability, $\cP(a|b)$ of producing a hardest microjet of flavour
$a$, given an initial parton of flavour $b$, is
\begin{equation}
  \label{eq:flavchange}
  \cP(a|b)=\int_0^1\,dz\, f^\text{hardest}_{a/b}(z)\,.
\end{equation}
The two main cases of interest are 
\begin{multline}\label{eq:flavchange-Q}
  \cP(g|q)
  = C_F t \left(\ln 4 - \frac{5}{8}\right)
  + \frac{t^2}{2}\big(-0.610848 C_A C_F 
      + 0.0519619 C_F^2 
      - 0.50753 C_F n_f T_R\big)\, + 
  \\
  % values with seed = 1,...,10
  + \frac{t^3}{6}\big( - 1.112(6) C_F^3 
      + 0.505(8) C_A^2 C_F 
      + 0.92(2) C_A C_F^2 
      + 0.89157(6) C_A C_F n_f T_R -
      \\
      - 0.0903(1) C_F^2 n_f T_R 
      + 0.338360(6) C_F n_f^2 T_R^2 \big)
  + \mathcal{O}(t^4)\,,
  % values with seed = 1
  % + \frac{t^3}{6}(-1.09751 C_F^3
  %     + 0.506414 C_A^2 C_F
  %     + 0.930485 C_A C_F^2
  %     + 0.891805 C_A C_F n_f T_R 
  %     \\ 
  %     - 0.089795 C_F^2 n_f T_R 
  %     + 0.33838 C_F n_f^2 T_R^2)
\end{multline}
and 
\begin{multline}\label{eq:flavchange-G}
  \cP(q|g) = \frac{2}{3}n_f T_R t 
  + \frac{t^2}{2}\big(-0.201036 C_A n_f T_R 
      - 0.438979 C_F n_f T_R 
      - 0.444444 n_f^2 T_R^2 \big)\, + 
  \\
  % values with seed = 1,..,10
  + \frac{t^3}{6}\big( 1.0498(2) C_A^2 n_f T_R 
      - 0.10560(6) C_A C_F n_f T_R 
      - 0.2537(1) C_F^2 n_f T_R +
      \\
      + 0.536081(8) C_A n_f^2 T_R^2 
      + 0.585304(6) C_F n_f^2 T_R^2 
      + 0.2962952(5) n_f^3 T_R^3 \big)
  + \mathcal{O}(t^4)\,.
  % values with seed = 1
  % + \frac{t^3}{6}(1.04976 C_A^2 n_f T_R 
  %     - 0.10544 C_A C_F n_f T_R 
  %     - 0.253678 C_F^2 n_f T_R + 
  %     \\
  %     + 0.536059 C_A n_f^2 T_R^2
  %     + 0.585293 C_F n_f^2 T_R^2
  %     + 0.296295 n_f^3 T_R^3)
\end{multline}
Analytical expressions for the $t^2$ coefficients are given in
Appendix~\ref{sec:app-jet-flav}.
The all-order results are shown in Fig.~\ref{fig:flavchange}.
One sees for example that for $R=0.4$, a quark (gluon) has a $4\%$
($6\%$) probability of becoming a differently flavoured microjet; for
$R=0.2$ the corresponding numbers are $7\%$ ($12\%$).
These numbers are subject to substantial higher-order (and jet- and
flavour-definition related) uncertainties, but they give an order of
magnitude for the maximal jet-flavour purity that can be obtained in
samples generated from flavour-pure partonic samples.

There are other questions that may also be interesting to ask about
jet flavour.
For example one might investigate it also in the context of inclusive
jet measurements, to identify how the flavour composition in a
steeply-falling hard-scattering spectrum is modified in the resulting
microjet spectrum.
It would also be conceptually interesting (though perhaps not very
physically relevant) to investigate the limit of asymptotically large
$t$, where the ratio of quark to gluon microjets might be expected to
tend to a fixed constant, independently of the whether the initial
parton is a quark or gluon.

%----------------------------------------------------------------------
\subsection{Multi (sub)jet observables}
\label{sec:multijet}

A number of methods developed for boosted electroweak and top-quark
tagging naturally involve small-radius subjets, as discussed in
various reviews~\cite{Boost2010,Boost2011,Boost2012,PlehnSpannowsky}.
These are used in part because the boost collimates the
heavy-object decay products and associated radiation and in part to
mitigate the impact of the very substantial pileup that is present at
the LHC. 

Here we will consider general purpose ``grooming'' approaches,
filtering~\cite{Butterworth:2008iy} and trimming~\cite{Krohn:2009th},
which, for their general ability to remove pileup, could have
applications also beyond boosted-objected studies.
In particular, it is interesting to ask how well filtered and trimmed
jets maintain the momentum of an original parton.
Similar considerations may apply also to the idea of building large jets
from small jets~\cite{Nachman:2014kla}.

%......................................................................
\subsubsection{Filtering}
\label{sec:filtering}

In filtering, one takes a jet clustered with an initial radius $R_0$,
reclusters its constituents on a smaller angular scale,
$R_\text{filt}<R_0$, and then discards all but the $n$ hardest
subjets.
Whereas $t$ in the previous sections was defined as being
$\frac{\as}{2\pi}\ln \frac{1}{R^2}$, plus higher orders from the
running coupling, we now imagine taking a large-radius original jet,
$R_0 = \order{1}$ and processing it with a small filtering radius,
with $t$ defined in terms of the filtering radius, $t \simeq
\frac{\as}{2\pi}\ln \frac{1}{R_\text{filt}^2}$, again plus higher
orders from the running coupling.
More generally, i.e.\ also for small $R_0$, $t \simeq
\frac{\as}{2\pi}\ln \frac{R_0^2}{R_\text{filt}^2}$ plus higher orders, and
the quantities we work out here will then relate the filtered jet to the
original jet rather than to the original parton.

We define $f^{k\text{-hardest}}(z)$ to be the probability that the
$k$-th hardest subjet carries a momentum fraction $z$ of the initial
parton (or large-$R$ jet). We can then express the energy loss between
the filtered jet and the initial parton as
\begin{equation}
  \label{eq:filtering}
  \langle\Delta z\rangle^{\text{filt},n} =
  \Bigg[\sum_{k=1}^{n}\int\,dz\,z\,f^{k\text{-hardest}}(z)\Bigg]-1.
\end{equation}
The total energy loss when taking the sum of the $n=2$ hardest microjets
is, for the case of an initiating quark
\begin{multline}\label{eq:filtpt-Q}
  % \langle\Delta p_t \rangle_q / p_t 
  \langle\Delta z \rangle^{\text{filt},2}_q 
  = \frac{t^2}{2}\big(-1.152 C_A C_F
      - 3.15229 C_F^2
      - 0.175607 C_F n_f T_R
  \big)\, +
  \\
  % values with seed = 1,...,10
  + \frac{t^3}{6}\big( 24.23(3) C_F^3 
      + 0.82448(2) C_A^2 C_F 
      + 6.2567(2) C_A C_F^2 
      + 0.893365(6) C_A C_F n_f T_R + 
      \\
      + 0.30444(2) C_F^2 n_f T_R 
      + 0.1170718(9) C_F n_f^2 T_R^2 \big)
  + \mathcal{O}(t^4)\,,
  % values with seed = 1
  % + \frac{t^3}{6}(24.3481403525  C_F^3 
  %     + 0.824448 C_A^2 C_F
  %     + 6.25577 C_A C_F^2 
  %     + 0.893344 C_A C_F n_f T_R
  %     \\
  %     + 0.304361 C_F^2 n_f T_R
  %     + 0.117069 C_F n_f^2 T_R^2)
\end{multline}
while for an initial gluon, we find
\begin{multline}\label{eq:filtpt-G}
  \langle\Delta z \rangle^{\text{filt},2}_g
  = \frac{t^2}{2}\big(- 3.88794 C_A^2
      - 0.5029C_A n_f T_R - 0.505401 C_F T_R n_f
  \big)\, +
  \\
  % values with seed = 1,...,10
  + \frac{t^3}{6}\big( 27.258(8) C_A^3 
      + 8.7362(2) C_A^2 n_f T_R 
      - 0.59419(2) C_A C_F n_f T_R +
      \\
      + 0.72083(2) C_F^2 n_f T_R 
      + 0.740071(4) C_A n_f^2 T_R^2 
      + 0.264690(3) C_F n_f^2 T_R^2 \big)
  + \mathcal{O}(t^4)\,.
  % values with seed = 1
  % + \frac{t^3}{6}(27.2537 C_A^3 
  %     + 8.73592 C_A^2 n_f T_R 
  %     - 0.594207 C_A C_F n_f T_R +
  %     \\
  %     + 0.720914 C_F^2 n_f T_R
  %     + 0.740068 C_A n_f^2 T_R^2
  %     + 0.264693 C_F n_f^2 T_R^2)
\end{multline}
The full analytical results for the coefficients of $t^2$ are given in
Appendix~\ref{sec:app-filtering}. 
If we consider the $n=3$ hardest subjets, then the first non-vanishing term is the
$t^3$ coefficient, and we find, for an initial quark
\begin{multline}
  \langle \Delta z\rangle^\text{filt,3}_q
  % values with seed = 1,...,10
  = \frac{t^3}{6}\big(-17.02(3) C_F^3 
      - 5.0299(2) C_A^2 C_F 
      - 18.9494(9) C_A C_F^2 
      \\
      - 0.643983(4) C_A C_F n_f T_R 
      - 3.1521(1) C_F^2 n_f T_R 
  \big)
  + \frac{t^4}{24}\, 8795(14) + \order{t^5}\,,
\end{multline}
and for an initial gluon
\begin{multline}
  \langle \Delta z\rangle^\text{filt,3}_g
  % values with seed = 1,...,10
  = \frac{t^3}{6}\big(-36.657(8) C_A^3 
      - 5.3268(2) C_A^2 n_f T_R 
      - 3.48461(7) C_A C_F n_f T_R 
      \\
      - 3.5362(3) C_F^2 n_f T_R 
      - 0.1526421(5) C_A n_f^2 T_R^2 
      - 0.1021532(8) C_F n_f^2 T_R^2
  \big) 
  \\
  + \frac{t^4}{24}\,79258(213) + \order{t^5}\,,
\end{multline}
where for the $t^4$ terms we have not extracted the explicit colour
separation.

\begin{figure}[pt]
  \centering
  \includegraphics[width=0.5\textwidth]{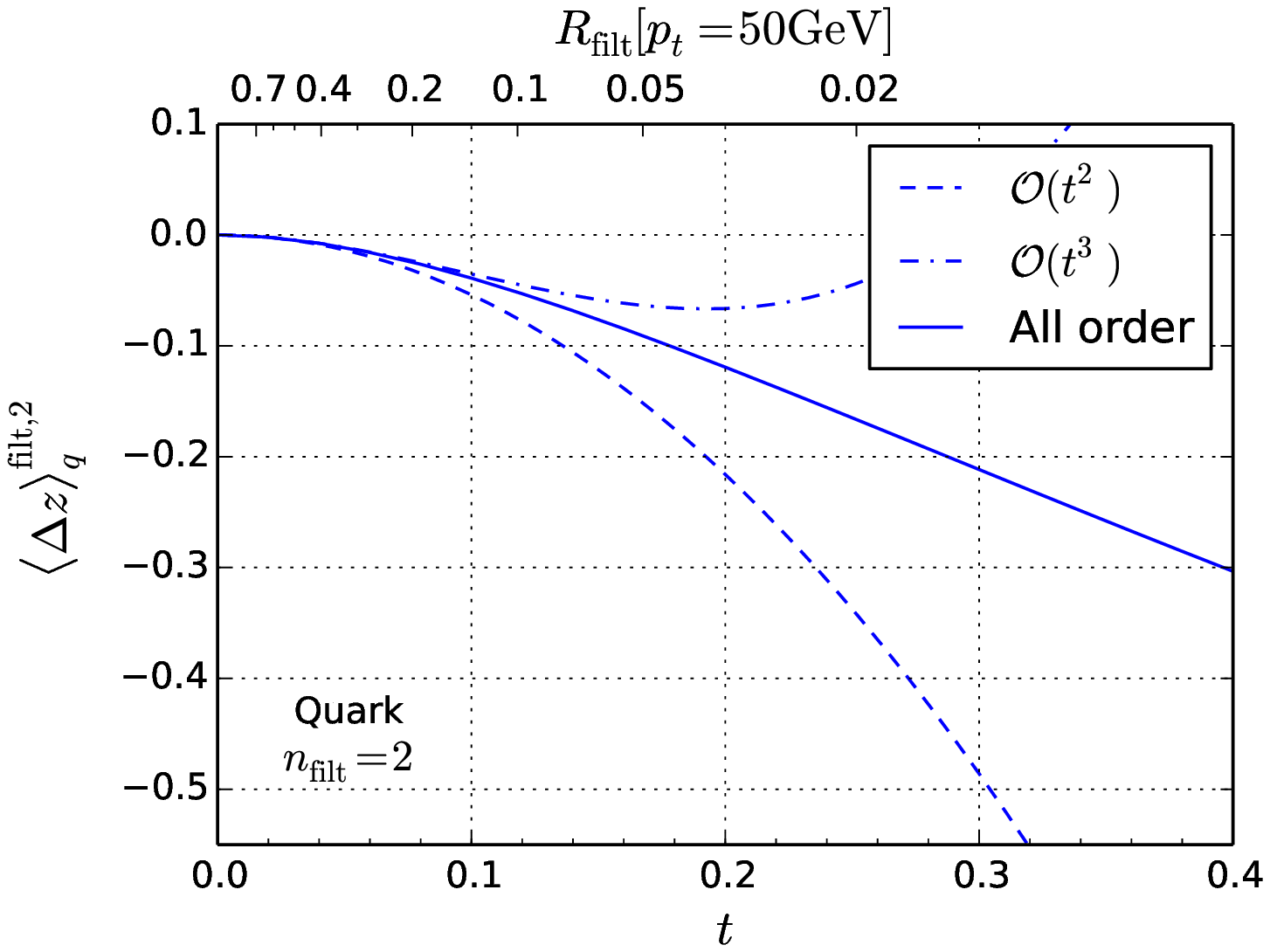}%
  \includegraphics[width=0.5\textwidth]{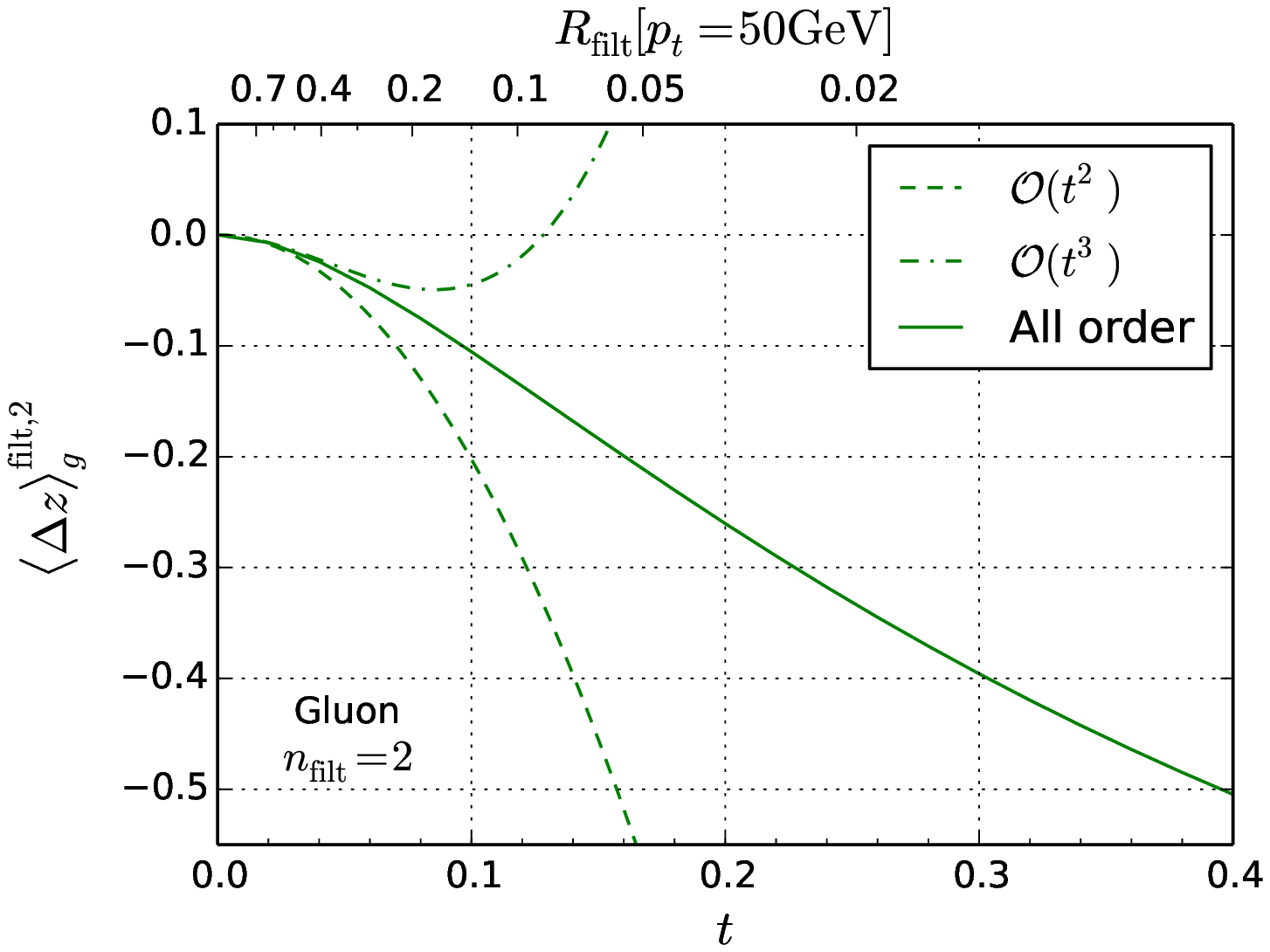}%
  \\
  \includegraphics[width=0.5\textwidth]{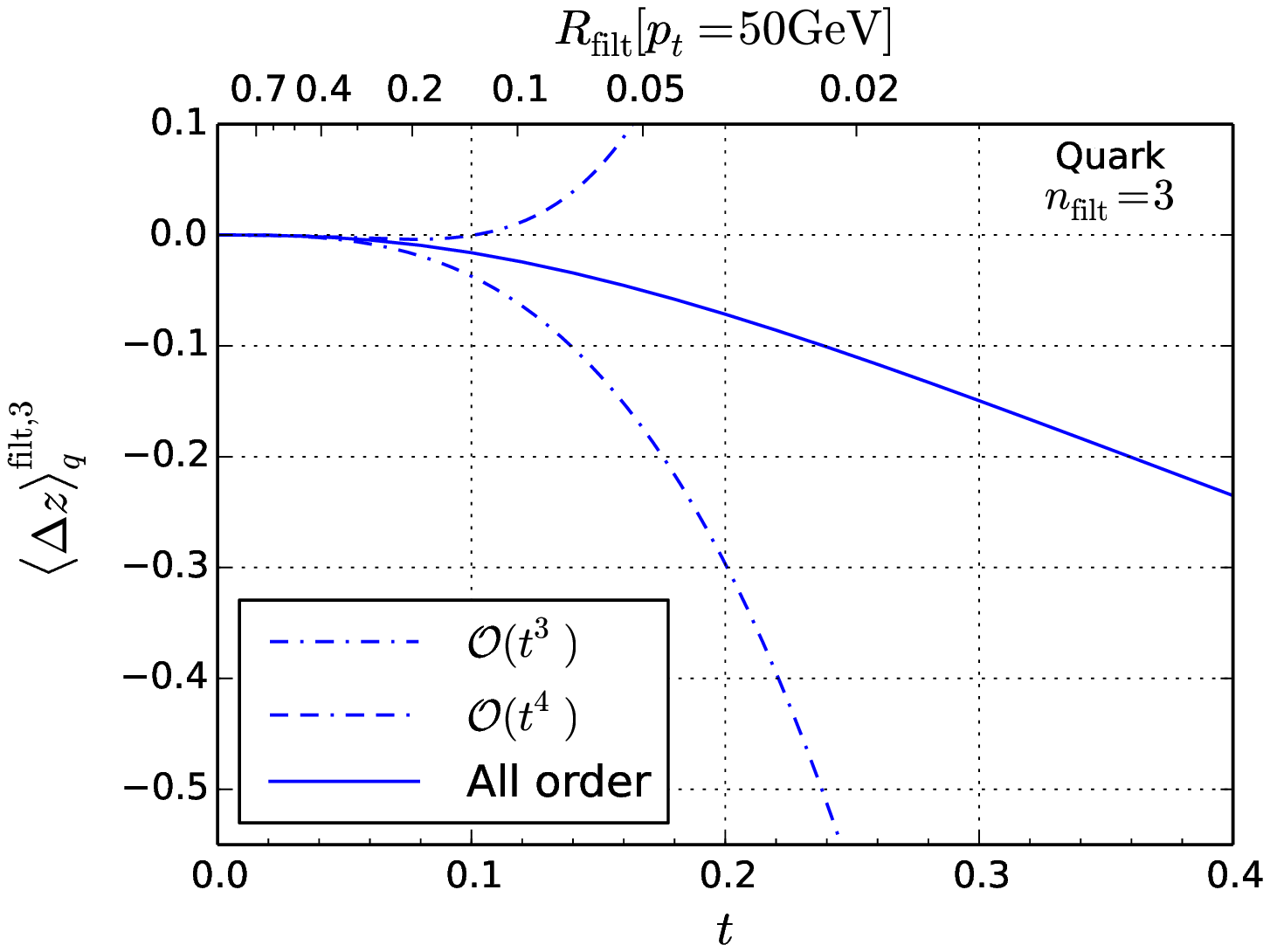}%
  \includegraphics[width=0.5\textwidth]{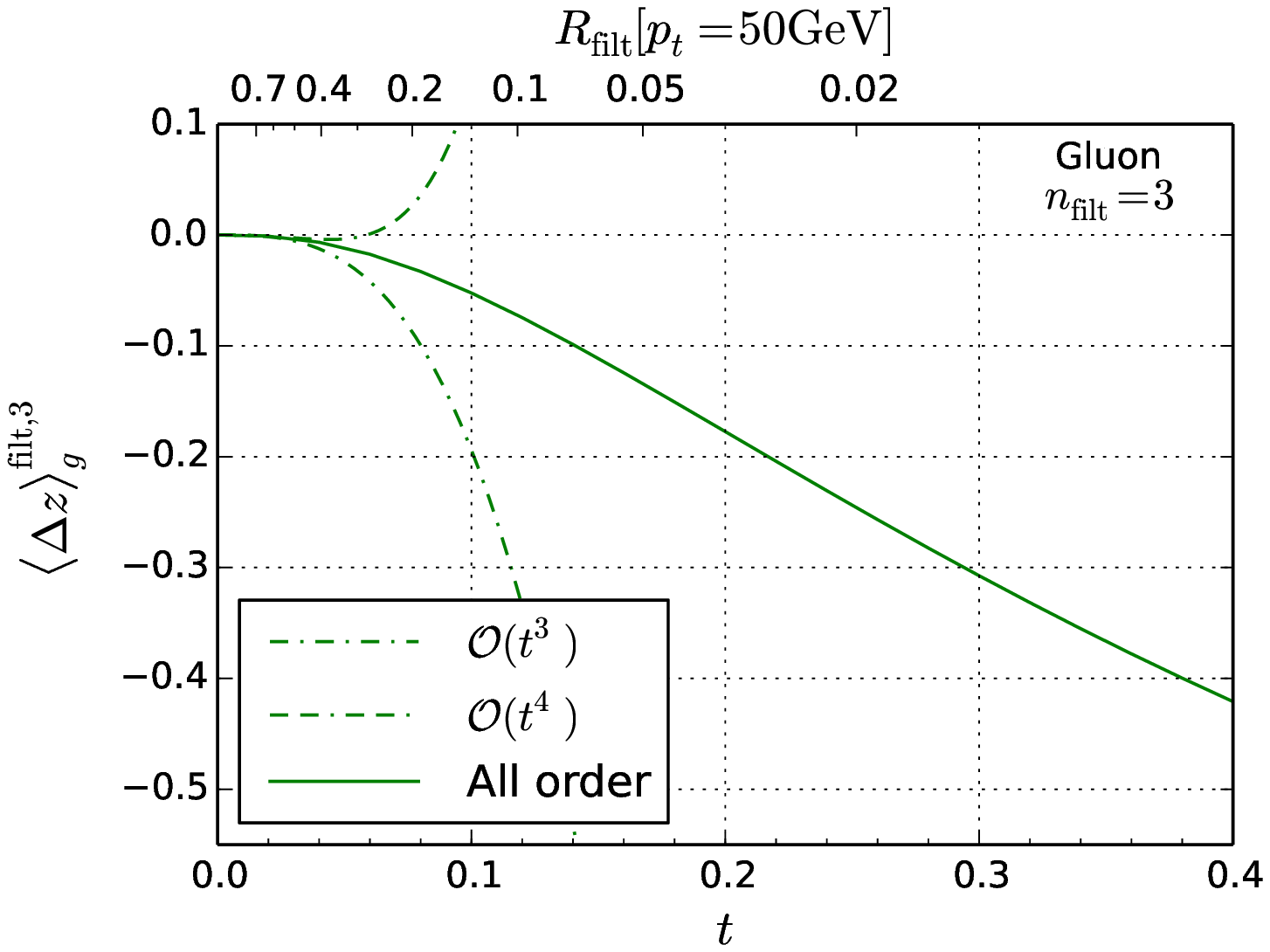}%
  \caption{Average fractional jet energy loss $\Delta z$ after filtering with
    $\nfilt=2$ (upper row) and $\nfilt=3$ (lower row), as a function
    of $t$, for quark-induced jets (left) and gluon-induced jets
    (right). Resummed results are represented as 
    solid lines. The second, third and fourth orders in $t$ are represented as
    dashed,  dash--dotted and dash--dash--dotted lines respectively.
    \label{fig:filtpt}
  }
\end{figure}

The all-order results for $\langle \Delta z\rangle^\text{filt}$ are
given in Fig.~\ref{fig:filtpt} for $\nfilt=2$ (upper row) and $\nfilt=3$ (lower
row). 
On one hand one observes how stable the filtered momentum is relative
to the momentum of a single microjet. 
Taking $t=0.1$, which corresponds to $R_\text{(filt)} \simeq 0.13$, an $n=2$
filtered jet retains $90\%$ of the parton's momentum (gluon case),
while an $n=3$ filtered jet retains nearly $95\%$ of the momentum. 
This is to be compared to just $75\%$ for a single microjet.
This is in part a consequence of the fact that $\langle \Delta
z\rangle^\text{filt,n}$ is non-zero only starting from order $t^{n}$
rather than order $t$.
Interestingly, however, the convergence of the series in $t$ seems to
be far worse. Considering the gluonic $n=3$ case, for $R_\text{filt}=0.2$ (a not
unusual choice), the zeroth order approximation is closer to the full
resummed result than any of the non-zero fixed-order results.
We are not sure why this is the case, but it suggests that resummation
effects should certainly be studied further if one is to carry out
precision physics with filtered jets beyond their originally intended
application of jet-mass determination in boosted-object taggers.

%......................................................................
\subsubsection{Trimming}
\label{sec:trimming}

In trimming one takes a jet of size $R_0$, reclusters its
constituents on a smaller angular scale $R_\text{trim} < R_0$ and keeps
(and merges) just the subjets with $p_t^\text{subjet} \ge \fcut
p_t^\text{jet}$.
As was the case for filtering, $t$ is now defined in terms of
$R_\text{trim}$ rather than $R$.

The energy difference between the trimmed jet and the initial parton of
flavour $i$ can then be expressed as a function of $\fcut$ by the
equation
\begin{equation}
  \label{eq:trimming}
  \langle\Delta z(\fcut)\rangle^\text{trim}_i =
  \Bigg[\sum_j\int_{f_{\rm cut}}^1 dz\,z\,f^{\rm incl}_{j/i}(z,t)\Bigg]-1=
  \sum_j\int_0^{\fcut} dz\, (-z)\, f^{\text{incl}}_{j/i}(z,t)\,.
\end{equation}
Given that the integral in the rightmost expression ranges only from
$z=0$ to $\fcut$, it is straightforward to see the that the result
will be of order $\fcut$.
For the quark case we have
\begin{multline}
  \label{eq:trimpt-Q}
  \langle\Delta z(\fcut)\rangle^\text{trim}_q
  =\frac{t}{2} C_F \Big[ 
      3 \fcut^2+4 \ln (1-\fcut)
  \Big]+
  \\
  % CF^2
  + \frac{t^2}{2}\Big[
   C_F^2 \fcut^2 \Big(\frac{3}{4} -
    \frac{1}{2} \ln \fcut \Big) 
  % CACF
  +C_A C_F \fcut \Big (4 \ln \fcut + \frac{8}{3} \Big)
  +C_A C_F \fcut^2 \Big(2 \ln \fcut-\frac{7}{6} \Big)
  \\
  % nf CF TR
  - \frac{4}{3} n_f C_F T_R \fcut 
  + n_f C_F T_R \fcut^2 \Big(-2 \ln \fcut - \frac{2}{3} \Big)
+ \mathcal{O}(\fcut^3)\Big]\,,
\end{multline}
where at order $t^2$ we have given just the first couple of terms in a
series expansion in $\fcut$ (see Appendix~\ref{sec:app-trimming} for
the full expressions).
For the gluon case we find
\begin{multline}
  \label{eq:trimpt-G}
  \langle\Delta z(\fcut)\rangle^\text{trim}_g = \\
  \frac{t}{6}\Big[ 
      C_A \fcut^2(12-4\fcut+3\fcut^2)
      - 2 T_R n_f \fcut^2 (3-4\fcut+3 \fcut^2)
      +12 C_A \ln (1-\fcut)
  \Big] +
  \\
  + \frac{t^2}{2}\Big[
  % CA^2
  C_A^2 \fcut \Big(4 \ln \fcut +\frac{10}{3}\Big)
  +C_A^2 \fcut^2\Big (2 \ln \fcut + \frac{2}{3}\Big)
  % nf CA TR
  - n_f C_A T_R \fcut^2 \Big (\frac{19}{6} + 2\ln \fcut \Big)
  \\
  % nf CF TR
  -\frac{8}{3} n_f C_F T_R \fcut
  - n_f C_F T_R \fcut^2 \ln \fcut
  % nf^2 TR^2
  + \frac{2}{3} \fcut^2 n_f^2 T_R^2
  + \mathcal{O}(\fcut^3)\Big]\,.
\end{multline}
A key feature of the above equations is the presence at order $t^2$ of
$\fcut \ln \fcut$ terms in the $C_F C_A$ and $C_A^2$ colour channels
for quarks and gluons respectively. 
This should not be surprising, since trimming has two small parameters,
$R_\text{trim}$ and $\fcut$, and to obtain a robust prediction it is
advisable to resum the logarithms that arise from each of these small
parameters. 
Terms $\as^n \fcut \ln^n R^2 \ln^{n-1} \fcut$ are in fact implicitly
included in our current approach, but ideally one would aim to resum
also all the terms that appear without $\ln R^2$ enhancements as well,
e.g.\ $\as^n \fcut \ln^{n-1} \fcut$.
The logarithms of $\fcut$ involve, for example, running-coupling
effects beyond those considered here, non-global logarithms~\cite{Dasgupta:2001sh},
and potentially also clustering logarithms~\cite{Delenda:2006nf}.
We leave their study to future work.
Without them, it seems difficult to use just the $\ln R^2$ resummation
to draw conclusions about the convergence and size of trimming
effects.
We therefore give our full $\ln R^2$-resummed results for trimming only
in Appendix~\ref{sec:app-trimming}, together with the complete
analytical expressions for the $t^2$ terms and numerical
determinations of the $t^3$ and $t^4$ contributions.

%======================================================================
\section{Conclusions}

In this article we have introduced a method to resum terms $(\as \ln
R^2)^n$ to all orders and applied it to a wide range of jet
observables.
One key observation is that small-$R$ effects can be substantial, for
example reducing the inclusive-jet spectrum by $30-50\%$ for gluon
jets when $R$ is in the range $0.4-0.2$.

The set of observables we considered should not be seen as exhaustive:
for example we might equally well have considered quantities such as
dijet asymmetries, microjet multiplicities or, with a little further
work, jet shapes.
The program that we used to obtain the results presented here can be
applied also to a number of these other observables, and will be made
public in due course to facilitate such studies.

One should not forget that the leading logarithms of $R$ that we
resummed are also included in parton-shower Monte Carlo programs that
include angular ordering, whether directly, or through some other
implementation of colour coherence.
However it is non-trivial to understand their role separately from the
many other physical effects present in Monte Carlo generators and it
is also difficult to merge them with other logarithmic resummations or
fixed-order calculations, especially beyond NLO.
With the resummations in the form that we have given here, one can
directly isolate the small-$R$ effects and merging with other
calculations becomes straightforward, as we illustrated briefly for jet
veto resummations. 
We also saw that the resummation made it possible to estimate the
range of validity in $R$ of fixed-order perturbative calculations.
One potential area of application, where $R$ is genuinely small, is in
substructure studies, where we saw for filtering that the resummation
effects are substantial, and fixed-order convergence intriguingly
poor.

We have not considered the detailed phenomenology of small-$R$ effects
in this paper, leaving such a study instead to future
work~\cite{smallR-pheno}.
This study should, among other things, also consider the role of
power suppressed  hadronisation corrections proportional to
$\Lambda_{\text{QCD}}/p_t$, which are known to receive a $1/R$
enhancement at small $R$~\cite{Korchemsky:1994is,Seymour:1997kj,Dasgupta:2007wa}.
Specific observables that could be examined are inclusive jet
spectra, jet-veto efficiencies in Higgs production and jet $p_t$
imbalances in heavy-ion collisions.
% 
% for such a study is the ratio of the inclusive jet
% cross-sections obtained with two different radius, as measured by the
% ALICE~\cite{Abelev:2013fn} and CMS~\cite{Chatrchyan:2014gia}
% collaborations and computed at NLO in \cite{Soyez:2011np}, and of
% course the jet-veto efficiency in Higgs production.
%
Other future work might consider small-$R$ logarithms in conjunction
also with threshold resummation.
Finally it would be of interest to understand how to go beyond LL
accuracy for small-$R$ resummations.
We believe that the techniques that apply would be of relevance also
for the resummation of jet substructure observables such as the
modified Mass Drop Tagger's jet-mass distribution beyond the LL
accuracy obtained so far~\cite{Dasgupta:2013ihk}.

%----------------------------------------------------------------------
\section*{Acknowledgements}

We are grateful to Matteo Cacciari for collaboration in the early
stages of this work and numerous discussions throughout,
to Andrea Banfi, Pier Monni and Giulia Zanderighi for helpful
suggestions, to Simone Alioli and Jon Walsh for exchanges about
their calculation of the jet veto efficiency
and to University College London for hospitality while this work was
being completed.
This work was supported by the European Commission under ITN grant
LHCPhenoNet, PITN-GA-2010-264564, by ERC advanced grant Higgs@LHC, by
the UK's STFC, by the French Agence Nationale de la Recherche, under
grant ANR-10-CEXC-009-01,
and by the ILP LABEX (ANR-10-LABX-63) supported by French state funds
managed by the ANR within the Investissements d'Avenir programme under
reference ANR-11-IDEX-0004-02.

%==============================================================

\appendix

\section{Analytical expressions}
\label{sec:appendix-analytic}

For all the second order coefficients given numerically in the main
text, this appendix provides the full analytical expressions. 
It also gives analytical and numerical results for some quantities
that for brevity were left out of the main text.

The results frequently involve the polylogarithm
$\mathrm{Li}_s(z)$, defined in the unit circle by 
\begin{equation}
  \label{eq:polylog}
  \mathrm{Li}_s(z)=\sum_{k=1}^\infty \frac{z^k}{k^s}\qquad  s\in\mathbb{N},\quad|z|<1\,,
\end{equation}
and by analytic continuation for $|z|>1$. The polylogarithm also
follows the recursive relation 
\begin{equation}
  \mathrm{Li}_1(z)=-\ln(1-z)\,, \qquad \mathrm{Li}_{s+1}(z)=\int_0^z\frac{\mathrm{Li}_s(t)}{t}dt\,.
\end{equation}

%......................................................................
\subsection[Hardest microjet ${\langle \Delta z \rangle}$]{Hardest microjet $\boldsymbol{\langle \Delta z \rangle}$}

\label{sec:app:avg-deltaz}
Let us first consider the energy difference between the the hardest
subjet and the initiating parton, as expressed in
Eq.~(\ref{eq:def-deltapt}). 
For the case of an initiating quark, given by Eq.~(\ref{eq:deltapt-Q}),
the $c_2$ coefficient of $\langle \Delta z\rangle^\text{hardest}$ is
\begin{multline}
  c_2[\langle \Delta z \rangle^\text{hardest}_q] =
  % CF^2 term
  C_F^2 \bigg[\frac{97}{144} - \pi^2 + \frac{5\ln2}{12} - 2\ln^2 2
      + 4\ln^2 3+8 \mathrm{Li_2}(\tfrac{2}{3})\bigg] +
  \\
  % CF nf term
  + 2 C_F T_R n_f \bigg[\frac{41}{12} + 12\ln 3 - 24\ln 2 \bigg] +
  \\
  % CF CA term
  + C_F C_A \bigg[ -\frac{967}{144}+\frac{\pi^2}{3}
      - 4\mathrm{Li_2}( -\tfrac{1}{3})
      - 4\mathrm{Li_2}(\tfrac{2}{3}) - 2 \ln^2 2\,-
      \\
      - 4\ln^23 + \frac{553}{12} \ln 2
      - \frac{99}{4} \ln 3 +8 \ln2 \ln 3 \bigg]\,,
\end{multline}
and when the initiating parton is a gluon, corresponding to
Eq.~(\ref{eq:deltapt-G}), we have
\begin{multline}
  c_2[\langle \Delta z \rangle^\text{hardest}_g] =
  % CA nf term
  C_A T_R n_f \bigg[\frac{3527}{432}-\frac{176}{3} \ln2
      + \frac{727}{24} \ln 3 \bigg]
  % nf^2 term
  + \frac{7}{72} {n_f^2 T_R^2} +
  \\
  % CF nf term
  + \frac{1}{432} C_F T_R n_f \bigg[451-3420 \ln2+1548 \ln 3 \bigg] +
  \\
  % CA^2 term
  + C_A^2\bigg[
    \frac{2 \pi ^2}{3}-\frac{2089}{288}-4 \ln ^2 2 - 2 \ln ^2 3 
    + \ln 2 \Big(\frac{1339}{24}+8 \ln 3\Big)-\frac{475}{16} \ln 3 -
    \\
    - 4 \text{Li}_2(\tfrac{2}{3})
    + 4 \text{Li}_2(\tfrac{1}{4})- 4 \text{Li}_2(\tfrac{3}{4})
    \bigg]\,.
\end{multline}

%......................................................................
\subsection[Logarithmic moment  ${\langle\ln z\rangle}$]{Logarithmic moment $\boldsymbol{\langle\ln z\rangle}$}

\label{sec:app-ln-pt}
Here we give our full set of results for the logarithmic moment of
$f^\text{hardest}$ as expressed in Eq.~(\ref{eq:def-logpt}).
Results for an initiating  quark were not given in detail in the main
text. 
The coefficients $c_{1\ldots3}$ of $t$,
$t^2/2$ and $t^3/6$ are respectively
\begin{subequations}
  \label{eq:logpt-Q}
\begin{equation}
  c_1[\langle \ln z \rangle^\text{hardest}_q]=
  \frac{C_F}{6} \big(9 - \pi^2 - 9 \ln 2 \big)\,,
\end{equation}
\begin{multline}
  c_2[\langle \ln z \rangle^\text{hardest}_q]=
  % CF CA term
  \frac{1}{432} C_F C_A \bigg [- 3887 + 786 \, \ln(\tfrac{9}{4})
    - 24 \pi^2 \Big(-92+54 \ln2 -30 \ln 3+\ln (\tfrac{729}{64})\Big)\, -
    \\
    - 144\Big(91 + 72\,\arccoth(5)\Big)\mathrm{Li_2}(\tfrac{1}{3}) 
    - 13392 \mathrm{Li_2}(\tfrac{2}{3})\,- 
    \\ 
    - 288 \ln^2 2 \Big(5+3 \ln 729 \Big)
    - 12 \ln 3 \Big(1175 + 24 \ln 9 (23+\ln 27)
     -72 \mathrm{Li_2}(\tfrac{1}{9}) \Big)\,+
    \\
    + 24 \ln 2 \Big(1166 - 33 \ln(\tfrac{9}{4})+36 \ln 3 (17+7 \ln 3) 
    + 72 \mathrm{Li_2}(-\tfrac{1}{2})
    + 144 \mathrm{Li_2} (\tfrac{2}{3})\Big)\,+
    \\
    + 288 \ln^3 2
    + 3456 \mathrm{Li_3}(-\tfrac{1}{2})
    + 1728 \mathrm{Li_3}(-\tfrac{1}{3})
    + 3456 \mathrm{Li_3} (\tfrac{2}{3})
    - 432 \zeta (3) \vphantom{\frac{1}{432}} \bigg] +
  \\
  % CF TR nf term
  + \frac{1}{216} C_F T_R n_f \bigg[1457 - 24\pi^2 
    + 36\ln 3 \Big(135 + \ln 81 \Big)
    - 12 \ln 2 \Big(833 + \ln 4096 \Big)
    + 288 \mathrm{Li_2}(\tfrac{2}{3}) \bigg] +
  \\
  % CF^2 term
  + \frac{1}{24} C_F^2\bigg[
    - 42 \text{Li}_2(-\tfrac{1}{2}) 
    - 126 \text{Li}_2(\tfrac{1}{3}) 
    - 30 \text{Li}_2(\tfrac{3}{4}) 
    - 96 \text{Li}_3(-\tfrac{1}{2}) 
    + 192 \text{Li}_3(\tfrac{1}{3}) 
    - 96 \text{Li}_3(\tfrac{2}{3})\, -
    \\
    - 24 \text{Li}_3(\tfrac{1}{9})
    - \ln 3 \Big(192 \text{Li}_2(\tfrac{2}{3}) 
    + 96 \text{Li}_2(-\tfrac{1}{3}) 
    - 102 + 128 \ln^2 3 + 63 \ln 3 \Big)\, +
    \\
    + 6 \ln 2 \Big(-16 \text{Li}_2(\tfrac{1}{3})
    + 16 \text{Li}_2(\tfrac{2}{3}) 
    + 16 \text{Li}_2(\tfrac{3}{4})
    - 41 + \ln 3 (31+24 \ln 3)\Big)\, -
    \\
    - 60 \zeta (3) + 9 + 208 \ln^3 2  
    - 3 (39+32 \ln 3) \ln^2 2 
    + 2 \pi^2 \Big(3+8 \ln 3\Big)\bigg]
  \\
  \simeq 
      - 0.73199 C_A C_F 
      + 1.0414 C_F^2 
      - 0.114427 C_F n_f T_R \,,
\end{multline}
and
\begin{multline}
    c_3[\langle \ln z \rangle^\text{hardest}_q]=
     - 0.17394(5) C_F^3 
    + 0.66631(2) C_A^2 C_F 
    - 0.81869(4) C_A C_F^2 
    +
    \\
    + 0.595241(9) C_A C_F n_f T_R\,
    - 0.534856(8) C_F^2 n_f T_R 
    + 0.076288(2) C_F n_f^2 T_R^2 \,.
\end{multline}
\end{subequations}
In the gluon case, expressed in Eq.~(\ref{eq:logpt-G}), the analytic
form for the second order coefficient is
\begin{multline}
  c_2[\langle \ln z \rangle^\text{hardest}_g]=
  % CA TR nf term
  C_A T_R n_f \frac{1}{216} \bigg[1697+24 \pi^2 +6336 \ln 3
    - 36 \ln 2 \Big(347+4 \ln 2\Big)\bigg] +
  \\
  % CF TR nf term
  + C_F T_R n_f \frac{1}{36} \bigg[97+8 \pi^2+48 \ln^2 2-48 \ln^2 3
    + 24 \ln (\tfrac{3}{2})+36 \ln (\tfrac{2187}{32768})
    - 96 \mathrm{Li_2} (\tfrac{2}{3})\bigg] -
  \\
  % CA^2 term
  - \frac{1}{144}C_A^2 \bigg[ 
    3624 \text{Li}_2(\tfrac{2}{3})
    + 8 \ln 2 \Big(
      108 \text{Li}_2(\tfrac{2}{3}) 
      - 1163 + 6 \ln 3 (9 \ln 3 - 22) 
      \Big)\,+
    \\
    + 12 \ln 3 \Big(
      48 \text{Li}_2(-\tfrac{1}{3}) 
      + 384 + \ln 3 (151 + 16 \ln 3)\Big)
    +36 \text{Li}_2(\tfrac{3}{4}) \Big(
      65 - 4 \ln (\tfrac{81}{8})\Big)\,-
    \\
    - 72 \zeta (3) + 1531 + 2208 \ln^3 2
    + 192 \ln^3(\tfrac{4}{3}) 
    - 2304 \ln^2 2 \Big(1+\ln (3)\Big) 
    + 288 \ln 3 \ln ^2(\tfrac{4}{3})\,+
    \\
    + 24 \pi^2 \Big(\ln (\tfrac{81}{64})-27\Big)
    + 12 \ln 4 \ln (\tfrac{4}{3}) \Big(151-60 \ln 2\Big)
  \bigg]
  \\
  + \frac{1}{54} {n_f^2 T_R^2} \big(23-24 \ln 2 \big)\,.
\end{multline}

%----------------------------------------------------------------------
\subsection{Jet flavour}
\label{sec:app-jet-flav}

The analytical form of the second-order coefficient for the
probability for an initiating quark to change flavour to a gluon
microjet, expressed numerically in Eq.~(\ref{eq:flavchange-Q}), is
\begin{multline}
  c_2[\cP(g|q)]=
  % CF nf Tr term
  -\frac{1}{12} C_F n_f T_R  \bigg[16 \ln 2-5\bigg] -
  \\
  % CF^2 term
  - \frac{1}{48} C_F^2 \bigg[-192\text{Li}_2(\tfrac{3}{4})
    + 16 \pi^2 +49+12 \ln 2 \Big(9-40 \ln 2\Big)
    + 48 \ln 3 \Big(\ln 256-3\Big)\bigg]+
  \\
  % CF CA term
  + \frac{1}{144} C_F C_A\bigg[288\text{Li}_2(\tfrac{3}{4})
    - 48\pi^2 +1463+5076\ln 3 \,+
    \\
    +36\ln 2\Big(-277+24\ln 2 -16\ln 3\Big)\bigg]\,.
\end{multline}
For the probability of a change of flavour from an initiating gluon to
a quark microjet jet,
corresponding to Eq.~(\ref{eq:flavchange-G}), we have 
\begin{multline}
  c_2[\cP(q|g)]=
  % CA nf Tr term
  - \frac{1}{27} C_A n_f T_R \bigg[455-2988 \ln 2+1476\ln 3\bigg]
  % nf^2 TR^2 term
  - \frac{12}{27} n_f^2 T_R^2 \,- 
  \\
  % CF Tr nf term
  - \frac{1}{27} C_F n_f T_R\bigg[288\ln(\tfrac{4}{3}) - 71\bigg]\,.
\end{multline}

%----------------------------------------------------------------------
\subsection{Filtering}
\label{sec:app-filtering}

We give next the analytical form of the $c_2$ coefficient for the
total energy loss when taking the sum of the 2 hardest microjets. For
an initial quark, corresponding to Eq.~(\ref{eq:filtpt-Q}), this is
\begin{multline}
  c_2[\langle \Delta z\rangle^\text{filt,2}_q]=
  % CF^2 term
  \frac{1}{18} C_F^2 \bigg[ 72 \text{Li}_2(\tfrac{3}{4})
  - 12 \pi ^2+13+9 \ln 2 \Big(3-32 \,\arccoth(5)\Big)\bigg] +
  \\
  % CF nf TR term
  + \frac{2}{9} C_F T_R n_f \bigg[43-216 \,\arccoth(5)\bigg]
  % CF CA term 
  + \frac{1}{12} C_A C_F \bigg[48 \text{Li}_2(\tfrac{2}{3})
    + 48 \text{Li}_2(-\tfrac{1}{3})-4 \pi ^2-111\,-
    \\
    - 96 \ln 2 (3+\ln 3)  + 3 \ln 3 \Big(99+16 \ln 3\Big)\bigg]\,,
\end{multline}
while for the gluon case, corresponding to Eq.~(\ref{eq:filtpt-G}),
we have 
\begin{multline}
  c_2[\langle \Delta z \rangle^\text{filt,2}_g]=
  % CA Tr nf term
  \frac{1}{216} C_A T_R n_f \bigg[2588+6480 \ln 2-6543 \ln 3\bigg]
  % CF Tr nf term
  + \\+ \frac{1}{108} C_F T_R n_f \bigg[146+324 \ln 2-387 \ln 3\bigg]\,-
  \\
  % CA^2 term
  - C_A^2 \frac{1}{48} \bigg[96 \text{Li}_2(\tfrac{1}{3})
    - 96 \text{Li}_2(\tfrac{2}{3})
    + 96 \text{Li}_2(\tfrac{1}{4})
    - 288 \text{Li}_2(\tfrac{3}{4})+
    \\
    + 48 \pi ^2 + 500 - 1425\ln 3 
    + 48\ln 2\Big(27-8 \ln 2+10 \ln 3\Big)\bigg]\,.
\end{multline}

%----------------------------------------------------------------------
\subsection{Trimming}
\label{sec:app-trimming}

\begin{figure}[tp]
  \centering
  \includegraphics[width=0.8\textwidth]{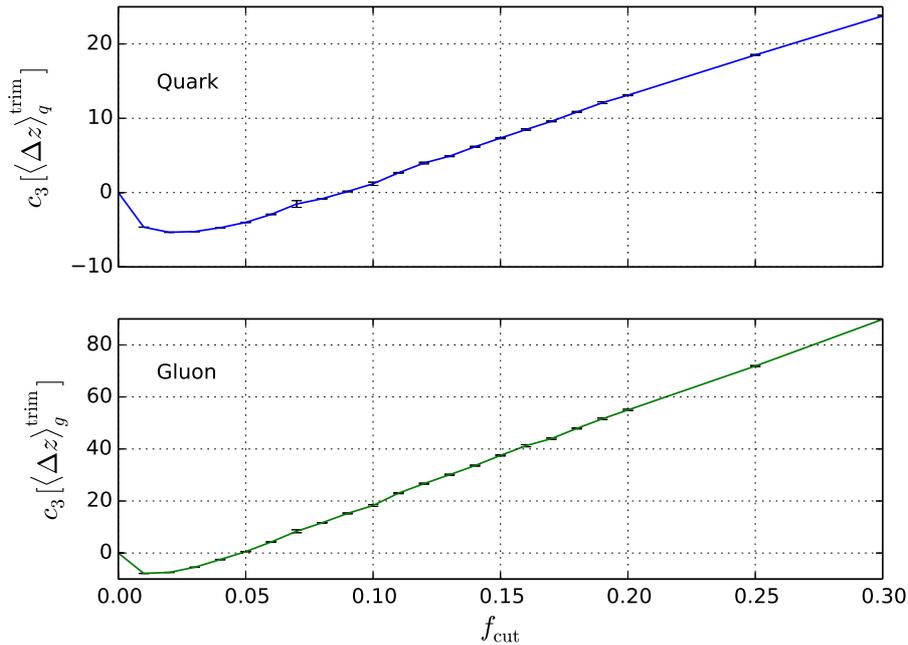}
  \caption{The third order coefficients $c_3(\langle \Delta
    z\rangle^\text{trim})$, as a function of $\fcut$, for quark (top) and
    gluon-induced (bottom) jets.
    The solid lines are
    simply intended to guide the eye and do not provide any
    information beyond what is specified by the points.
    \label{fig:trim-t3}
  }
\end{figure}

The full second order results for $\langle\Delta z\rangle^\text{trim}$
are
\begin{multline}
  c_2[\langle\Delta z\rangle^\text{trim}_q]=
  % CF^2 term
  \frac{1}{12} C_F^2 \bigg[-30 \fcut^2 \ln \fcut 
    + 16 \ln (1-\fcut) \Big(3 \fcut^2
    + 3 \ln (1-\fcut)-2\Big)\,+
    \\
    + 48 \text{Li}_2(1-\fcut)
    + 16 \Big(\fcut^2+6\Big) \fcut \arctanh(1-2 \fcut)
    + 6 \fcut^3+5\fcut^2+16 \fcut-8 \pi ^2\bigg]+
  \\
  % CA CF term
  + \frac{1}{6} C_A C_F \bigg[(\fcut-1) \Big(
    -(2 \fcut^3+9\fcut^2+32\fcut)
    - 4 ((\fcut-2) \fcut+4) \ln (1-\fcut)\Big)\,+
    \\
    + 4 \fcut \Big(2 \fcut^2+3\fcut+6\Big) \ln \fcut\bigg]+
  \\
  % CF nf TR term
  + \frac{2}{3} C_F T_R n_f \fcut \bigg[(\fcut-1) (\fcut+1)
    (\fcut+2)-\fcut (2 \fcut+3) \ln \fcut\bigg]\,,
\end{multline}
for the quark case, 
and 
\begin{multline}
  c_2[\langle\Delta z\rangle^\text{trim}_g]=
  % CA^2 term
  \frac{1}{36} C_A^2 \bigg[-36 (\fcut-4) \fcut^3 \ln \fcut
    + 24 \ln (1-\fcut) \Big((3\fcut^2-4\fcut+12) \fcut^2 +
    \\
    + 6 \ln (1-\fcut)\Big) + 144 \text{Li}_2(1-\fcut) 
    - 75 \fcut^4+ 16\fcut^3-84\fcut^2+264 \fcut-24 \pi ^2\bigg]-
  \\
  % CA Tr nf term
  - \frac{1}{18} C_A T_R n_f \bigg[
    12 \Big(3 \fcut^4-4\fcut^3+3\fcut^2+2\Big) \ln (1-\fcut)
    + 12 \fcut^2 \Big(8 \fcut+3\Big) \ln \fcut - \\
    - 57 \fcut^4 + 8\fcut^3+69\fcut^2+24\fcut\bigg]+
  \\
  % CF Tr nf term
  + \frac{1}{3} C_F T_R n_f \bigg[
    \Big(4-6 \fcut^2\Big) \ln (1-\fcut)
    + 4 \Big(4\fcut^3-3 \fcut^4\Big) \arctanh(1-2 \fcut)\,-
    \\
    - 3 \fcut^2 \ln \fcut + 2\fcut^4 + 2\fcut^2- 4\fcut \bigg]
  % TR^2 nf^2 term
  + \frac{2}{9} T_R^2 n_f^2 \fcut^2 \Big(3 \fcut^2-4\fcut+3\Big) \,,
\end{multline}
for the gluon case.
Numerical results for the third 
%and fourth 
order terms are given in
Figs.~\ref{fig:trim-t3}.

\begin{figure}[pt]
  \centering
  \includegraphics[width=0.5\textwidth]{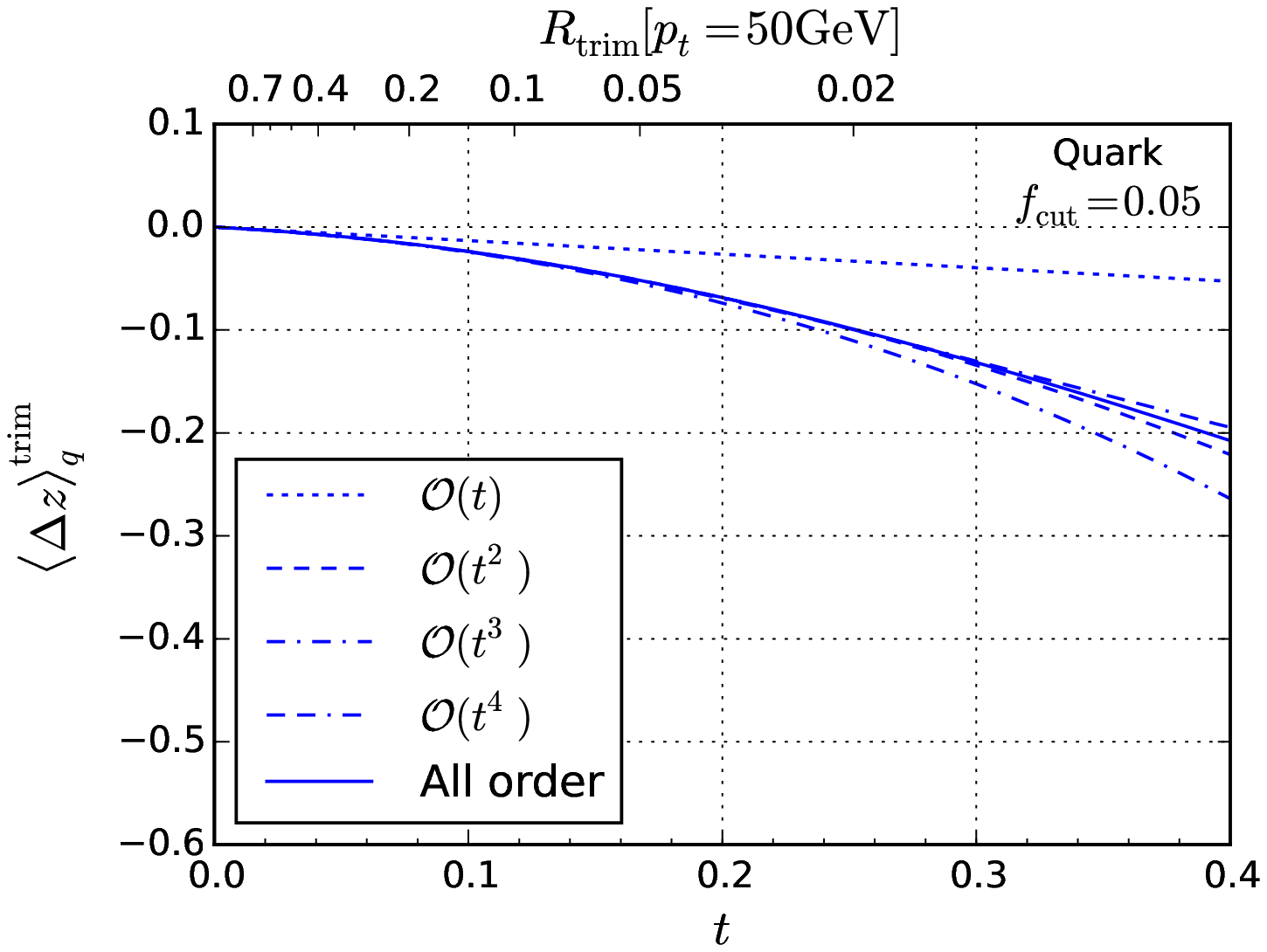}%
  \includegraphics[width=0.5\textwidth]{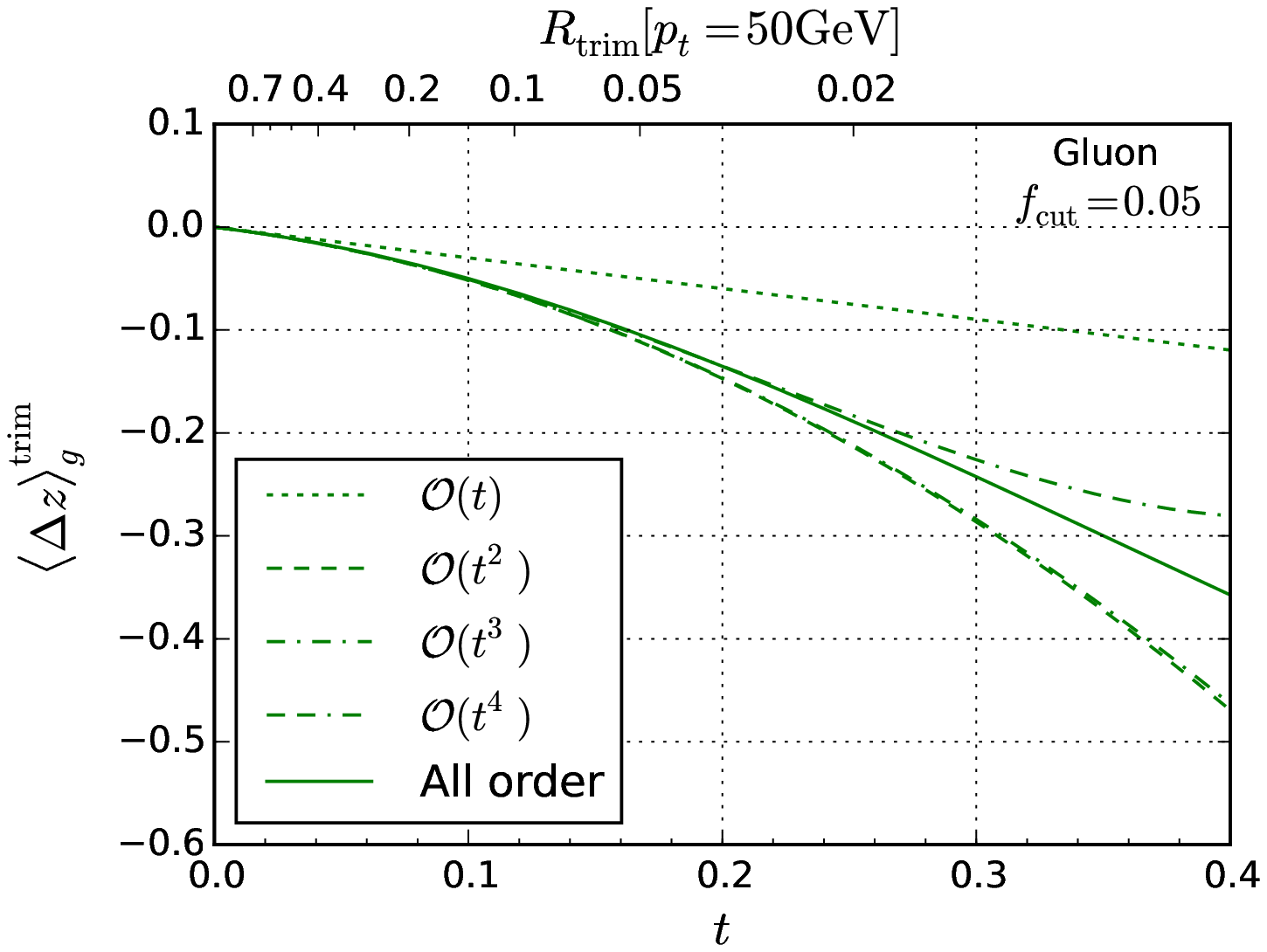}
  \caption{Average fractional jet energy loss $\Delta z$ after trimming with
    $\fcut=0.05$ as a function of $t$ for quarks (left) and gluons
    (right), with $\ln R^2_\text{trim}$ resummation, but not $\ln \fcut$
    resummation. 
    \label{fig:trimpt}
  }
\end{figure}

All-order results for the widely used choice of $\fcut=0.05$ are shown
in Fig.~\ref{fig:trimpt} as a function of $t$.
As with filtering, the energy loss from trimmed jets with a given
$R_\text{trim}$ is much reduced relative to that from a single
microjet with that same radius $R_\text{trim}$.
With this specific $\fcut$ value, the energy loss is somewhat smaller even
than for filtering. 
Additionally the convergence of the series appear to be far better.
However, as was discussed in section~\ref{sec:trimming}, one should
keep in mind that trimming also leads to logarithms of $\fcut$, only
some of which are included in our resummation.
Their potential importance can be appreciated by considering that
the full calculation would involve an integral over $\as$ down to
scale $\fcut R_\text{trim} p_t$, which is much smaller than the scale
$R_\text{trim} p_t$ that is included in our small-$R$ resummation.
Indeed for the choice of $p_t = 50 \GeV$ and $\fcut= 0.05$ that we use
in Fig.~\ref{fig:trimpt}, any $R_\text{trim} < 0.4$ would force us to
consider scales below $1\GeV$.

%======================================================================
\section{Comparisons and fixed-order cross checks}
\label{sec:comparisons-etc}

%----------------------------------------------------------------------
\subsection{Comparison with the literature}
\label{sec:alioli-walsh}

A numerical calculation of the $\as^3 \ln \frac{Q}{p_t} \ln^2
R^2$ contribution to the jet veto efficiency in the case of
$gg\to H$ production was given recently by Alioli and
Walsh (AW)~\cite{Alioli:2013hba}.
This section provides the detailed comparison.

First let us rewrite the first two orders of Eq.~(\ref{eq:logpt-G}) as
an expansion in powers of $\as$, taking into account
Eq.~(\ref{eq:t-power-series}) for the conversion between the $t$
and $\as$ expansions:
\begin{multline}
  \label{eq:lnzg-as-expansion}
  \langle \ln z \rangle _g^\text{hardest}= 
  \frac{\as(p_t)}{2\pi}\ln\frac{1}{R^2} \bigg[
      \frac{1}{72} C_A \left(131 - 12 \pi ^2 - 132\ln 2\right)
      + \frac{1}{36} n_f T_R (-23+24 \ln 2)
  \bigg]
  \\
  + \frac{\as^2(p_t) }{4 \pi^2}\ln^2\frac{1}{R^2}\left[
    - 0.36982 C_F n_f T_R
    + 0.117861n_f^2T_R^2
    + 0.589237 C_A n_f T_R 
    -0.901568 C_A^2
  \right] +\\+ \order{\as^3}\,,
\end{multline}
where $p_t$ is the transverse-momentum above which jets are vetoed.
AW have an expansion for the $R$-dependent terms in the jet-veto
probability, Eqs.~(1.1) and (1.3) of Ref.\cite{Alioli:2013hba}, whose
leading logarithms of $R$ read
\begin{equation}
  \label{eq:AlioliWalsh-notation}
  \exp\left[ C_n^{(n-1)}  \left(\frac{\as(p_t) C_A}{\pi}\right)^n \ln
    \frac{Q}{p_t}\ln^{n-1} R^2 \right]\,. 
\end{equation}
Substituting Eq.~(\ref{eq:lnzg-as-expansion}) into
Eq.~(\ref{eq:jet-veto-probability-microjet-processed}) gives our
result for $C_3^{(2)}$\,,
\begin{equation}
  \label{eq:as3_L_lnR2}
  C_3^{(2)} = \left(0.36982 \frac{C_F n_f T_R}{C_A^2} - 0.589237 \frac{n_f T_R}{C_A} 
    - 0.117861 \frac{n_f^2 T_R^2}{C_A^2} + 0.901568\right) 
  \simeq 0.46566\,.
\end{equation}
The results from AW that can be straightforwardly compared are the $C_F
n_f T_R$ term and the full result.
For the $C_F n_f T_R$ term, including the explicit colour factors, our
result is $0.36982 \frac{C_F n_f T_R}{C_A^2} \simeq 0.1370$ (for $n_f
= 5$), with which AW's result of $0.1405 \pm 0.0011$ is marginally
compatible.
For the full result, our coefficient of  $C_3^{(2)} \simeq 0.46566$ is
to be compared with that from AW of  $C_3^{(2)} =
-0.356\pm0.011$. 
After consulting with the authors, they identified the origin of the
difference as being due to the fact that they effectively did not
include the running coupling corrections that appear at order $\as^2$ in the
expansion of $t$.
Once this is resolved, their answers come into reasonable agreement
with ours.
%

%----------------------------------------------------------------------
\subsection{Fixed-order cross checks}
\label{sec:fixed-order-cross-checks}

On one hand the results we have presented are sufficiently
straightforward that there should be no need for cross-checks from
full fixed-order calculations.
However, given the initial disagreement with the results of
\cite{Alioli:2013hba}, such cross--checks became desirable.
In principle one could take a program such as MCFM and directly take
the limit of small $p_t$ and small $R$ in the Higgs plus two jet
process at NLO, in order to directly cross--check the $C_3^{(2)}$
coefficient in Eq.~(\ref{eq:as3_L_lnR2}).
However in practice, we believe that it would be almost impossible to
take the appropriate double limit, simultaneously satisfying the
requirements of being sufficiently asymptotic and having sufficient
Monte Carlo statistics.

Instead we have opted to test the basic framework and in particular
the results for the distribution $f^\text{hardest}(z)$ integrated up to
some finite $z$, 
\begin{equation}
  P_{z<z_{\max}} = \int_0^{z_{\max}} dz\, f^\text{hardest}(z)\,,
\end{equation}
which is closely related to (one minus) the
probability of vetoing a jet.
We examined the $e^+e^- \to 3$~jet process at next-to-leading order
(NLO), using the Event2 program~\cite{Catani:1996vz}.
We initially clustered the events with the $e^+e^-$ version of the
inclusive Cambridge/Aachen
algorithm~\cite{Dokshitzer:1997in,Wobisch:1998wt}, ``ee-genkt'' with
$p=0$ and $R_0=1.5$ as defined in FastJet~\cite{Cacciari:2011ma}.
Then, for each jet with an energy above $z_{\max} Q/2$,
where $Q$ is the centre-of-mass energy, we
progressively unclustered the jet until we reached a configuration
where each of the subjets had an energy of less than $z_{\max} Q/2$.
We identified the Cambridge/Aachen distance for that declustering,
\begin{equation}
  \label{eq:dij}
  d^\text{(C/A)} = \frac{1-\cos\theta}{1-\cos R_0} \simeq
  \frac{\theta^2}{2(1-\cos R_0)} \,,
\end{equation}
where $\theta$ is the angle between the two subjets or particles that
got declustered, and the rightmost (approximate) expression holds for small $\theta$.
We then added an entry for that event to the corresponding bin of $\ln
1/d^\text{(C/A)}$.
This gives us a numerical result for
\begin{equation}
  \label{eq:zmax-integral}
  \frac{d P_{z<z_{\max}}}{d\ln 1/d^\text{(C/A)}} 
  = \frac{d}{d\ln 1/d^\text{(C/A)}} 
  \int_0^{z_{\max}} dz f^\text{hardest}(z)\,,
\end{equation}
because for values of $d^\text{(C/A)}$ smaller than that of the bin,
the hardest $z$ satisfies $z > z_\text{max}$, while for values of
$d^\text{(C/A)}$ larger than that of the bin the hardest $z$ satisfies
$z < z_\text{max}$.

\begin{figure}
  \centering
  \includegraphics[width=0.48\textwidth]{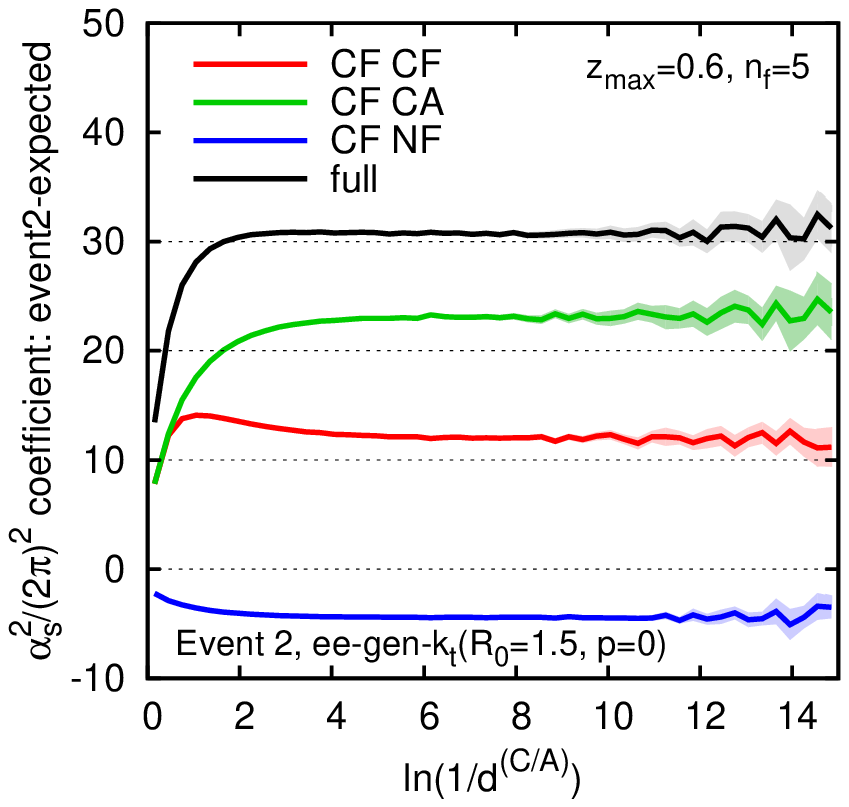}\hfill
  \includegraphics[width=0.48\textwidth]{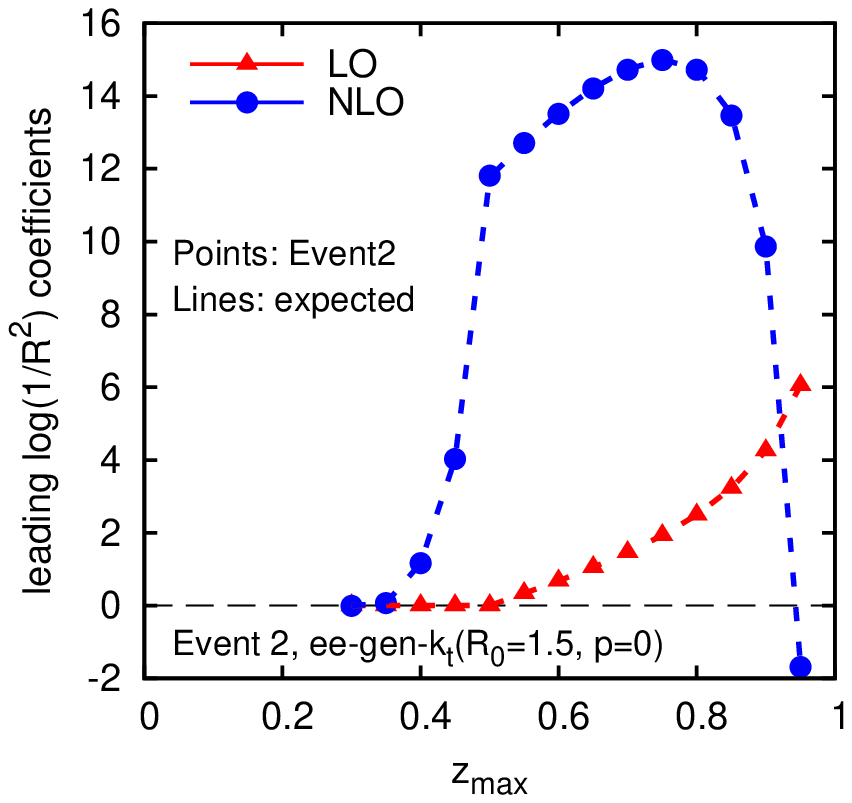}\hfill
  \caption{Left: difference between Event2 NLO results and the expected
    $\as^2 \ln 1/d^\text{(C/A})$ coefficient for the derivative with respect to
    $\ln 1/d^\text{(C/A)}$ of $P_{z<z_{\max}}$, for $z_{\max} = 0.6$. 
    The lines should be flat asymptotically if Event2 and our
    calculations agree.
    Right: comparison of the expected and Event2 results for the $[(\as/2\pi)
    \ln 1/R^2]^n$ coefficients for $P_{z<z_{\max}}$, as a function of
    $z_{\max}$, for $n=1$ (LO) and $n=2$ (NLO). 
  }
  \label{fig:fixed-order-checks}
\end{figure}

Up to order $\as^2$, we expect the result to be
\begin{multline}
  \label{eq:expected-cross-check}
  \frac{dP_{z<z_{\max}}}{d\ln 1/d^\text{(C/A)}}
  =
  \frac{\as}{2\pi} \left[c_1[P_{z<z_{\max}}] + \order{d^\text{(C/A)}} \right]
  +\\+
  \left( \frac{\as}{2\pi} \right)^2 \left[ 
    \big(b_0 c_1[P_{z<z_{\max}}] + c_2[P_{z<z_{\max}}] \big) \ln
    \frac{1}{d^\text{(C/A)}} 
    + \order{1}\right]
  + \order{\as^3}\,,
\end{multline}
where the $c_1$ and $c_2$ coefficients are precisely those worked out
in this paper at LL order in the small-$R$ limit (had we evaluated
subleading terms, care would have been needed concerning the
difference between $d^{(C/A)}$ and $R^2$, however since this just
reduces to a constant factor for small angles, it is irrelevant at LL
order).
Figure~\ref{fig:fixed-order-checks} (left) shows the difference
between the actual NLO term in Event2 (without the LO contribution)
and the expected $\order{\as^2}$ term of
Eq.~(\ref{eq:expected-cross-check}), as evaluated from the first and
second order expansions of our small-$R$ resummation.
It is shown as a function of $\ln 1/d^\text{(C/A)}$, for
$z_{\max}=0.6$, both for the total result and separately for each of
the colour-factor contributions.
In all cases asymptotically it is flat, indicating that we have
subtracted the correct $\order{\as^2 \ln \frac1{d^\text{(C/A)}}}$ contribution.
The right-hand plot shows the expected results for the coefficients of
the $\order{\as}$ and $\order{\as^2}$ LL terms, versus those observed
in Event2, as a function of $z_{\max}$.
Again the agreement is very good. 
We could also have directly studied $\langle \ln z
\rangle^\text{hardest}$, however the test as shown in
Fig.~\ref{fig:fixed-order-checks} is in some respects more complete
because it probes the $f^\text{hardest}(z)$ distribution
differentially.
Two final remarks are in order.
Firstly, these tests do not unequivocally demonstrate that it is the
$\langle \ln z \rangle^\text{hardest}$ moment that is the relevant
quantity for a jet-veto probability --- for this one still relies on
the calculation in
Eqs.~(\ref{eq:jet-veto-probability-partons-start})--(\ref{eq:jet-veto-probability-microjet-processed}).
Secondly, Event2 provides a test of our results only for quark
jets. 
Nevertheless, the property of angular ordering that we rely on for our
calculations is common to quark and to gluon jets.

%======================================================================
%======================================================================
%======================================================================
%======================================================================
%======================================================================
%======================================================================
%======================================================================
%======================================================================
%======================================================================
%======================================================================
%======================================================================
%======================================================================
%======================================================================
%======================================================================
%======================================================================

\end{document}